%% file: main.tex
\documentclass[twocolumn]{article}
\pdfoutput=1

\usepackage[table]{xcolor}
\usepackage{enumerate, amsfonts, amsmath, multirow, tikz, algpseudocode, xargs, subfig}
\usetikzlibrary{arrows,automata,calc,positioning}

\newtheorem{mydef}{Definition}

\newcommand{\red}[1]{#1}

\algnewcommand\algorithmicforeach{\textbf{for each}}
\algdef{S}[FOR]{ForEach}[1]{\algorithmicforeach\ #1\ \algorithmicdo}

\newcommandx{\knut}[2][1=]{\todo[linecolor=red,backgroundcolor=red!25,bordercolor=red,#1]{#2}} 
\newcommandx{\sajed}[2][1=]{\todo[linecolor=green,backgroundcolor=green!25,bordercolor=green,#1]{#2}} 
\newcommandx{\koen}[2][1=]{\todo[linecolor=blue,backgroundcolor=blue!25,bordercolor=blue,#1]{#2}} 
\newcommandx{\nick}[2][1=]{\todo[linecolor=yellow,backgroundcolor=yellow!25,bordercolor=yellow,#1]{#2}} 
\newcommandx{\alex}[2][1=]{\todo[linecolor=red,backgroundcolor=red!25,bordercolor=red,#1]{#2}} 

\begin{document}
%
\title{Enhancing Temporal Logic Falsification with Specification Transformation and Valued Booleans}
%
%
%
\date{}

\author{Johan~Lid\'{e}n~Eddeland \\Volvo Car Corporation\\johan.eddeland@volvocars.com\\johan.eddeland@chalmers.se \and
        Koen~Claessen \\Chalmers University of Technology\\koen@chalmers.se\and
        Nicholas~Smallbone \\Chalmers University of Technology\\nicsma@chalmers.se \and
        Zahra~Ramezani \\Chalmers University of Technology\\rzahra@chalmers.se \and
        Sajed~Miremadi \\Volvo Car Corporation\\sajed.miremadi@volvocars.com\and
        Knut~\AA{}kesson \\Chalmers University of Technology\\knut@chalmers.se}
\maketitle

\begin{abstract}
Cyber-Physical Systems (CPSs) are systems with both physical and software components, for example cars and industrial robots. Since these systems exhibit both discrete and continuous dynamics, they are complex and it is thus difficult to verify that they behave as expected. Falsification of temporal logic properties is an approach to find counterexamples to CPSs by means of simulation. In this paper, we propose two additions to enhance the capability of falsification and make it more viable in a large-scale industrial setting. The first addition is a framework for transforming specifications from a signal-based model into Signal Temporal Logic. The second addition is the use of Valued Booleans and an additive robust semantics in the falsification process. We evaluate the performance of the additive robust semantics on a set of benchmark models, and we can see that which semantics are preferable depend both on the model and on the specification.
\end{abstract}


%

\input{Sections/1_introduction.tex}

\input{Sections/2_falsification.tex}
\input{Sections/3_signal_based_specifications.tex}

\input{Sections/4_vbools.tex}
\input{Sections/5_experiments.tex}

\input{Sections/6_conclusion.tex}

\section*{Acknowledgment}
The authors would like to thank Alexandre Donz\'{e} for his helpful comments on a draft of this paper. This work has been performed with support from the Swedish Governmental Agency for Innovation Systems (VINNOVA) project TESTRON 2015-04893 and from the Swedish Research Council (VR) project SyTeC 2016-06204. This support is gratefully acknowledged.


\bibliographystyle{IEEEtran}
\bibliography{references}

\end{document}

%% file: Sections/1_introduction.tex
\section{Introduction}

Assuring the quality of Cyber-Physical Systems (CPSs) is an important task that is growing more and more complex. Industrial-size systems with both discrete and continuous dynamics, \emph{i.e.} hybrid systems, require durable methods for design automation \cite{seshia2017design}, as well as validation methods that are beyond the current capabilities of \emph{e.g.} model-checking \cite{clarke2009model}. Since the general problem of finding the set of reachable states for this kind of systems is undecidable \cite{henzinger1995s}, we instead resort to testing the systems. For testing and/or monitoring of CPSs, there are many possible approaches (see \cite{bartocci2018specification, kapinski2016simulation} for \red{two surveys}) -- in this work, we consider \emph{falsification} of temporal logic specifications. \red{Another approach is deductive methods for proving properties of CPSs \cite{platzer2018logical}, but in many industrial applications there is no mathematical model to analyze, instead there is only the possibility to simulate the system under test.} Falsification can be done for CPSs both with the actual hardware, or as in the case of this paper, where the hardware is being simulated. 

Falsification of temporal logic specifications for CPSs is a method which attempts to find counterexamples to properties of systems by optimization over \emph{robustness} of the specification. Here, robustness is a measure of distance to violation of the specification. The falsification framework has been shown to be useful for several different applications \cite{fainekos2012verification, annpureddy2011s}, and it can still be modified in many different ways. For example, one can consider different optimization algorithms to search for the counterexample (\emph{e.g.} ant colony optimization \cite{annapureddy2010ant} or functional gradient descent \cite{abbas2014functional}). 

Falsification requires use of a formal specification, typically written in \emph{Metric Interval Temporal Logic} (MITL) \cite{koymans1990specifying} or \emph{Signal Temporal Logic} (STL) \cite{maler2004monitoring} (or some variant thereof). However, these formal logics are not currently well established in industry, \red{since the specifications used in industry need to be understood by engineers from many different disciplines.} This means that it can be difficult to apply falsification when there is no formal specification available to test against. 

In an attempt to tackle this problem, we present a framework for transforming requirements modelled in a causal, signal-based language (\emph{e.g.} Simulink \cite{simulink}) into specifications in STL. This allows expert test engineers to model executable requirements using a tool they are familiar with, while also making falsification possible for the models under development. 

As an additional measure to enhance the falsification process for industrial-size problems, we apply an alternative robust semantics to be used in the falsification problem. Specifically, we use the additive semantics presented for the logical framework Valued Booleans \cite{claessen2018using}. We evaluate the performance of additive semantics for several specifications and see in which cases they are preferable to the ``standard'' semantics of STL robustness. \red{To be clear, these changes apply when we attempt to falsify a specification by means of optimization, rather than by performing brute-force exploration of system executions. }

\subsection{Related work}
The main focus of this paper is to adapt the framework of falsification to work better in certain industrial applications. The tools Breach \cite{DonzeBreach} and S-TaLiRo \cite{annpureddy2011s} are used to perform falsification with STL and MTL, respectively. Both of these tools are based on the idea of a robustness measure for temporal logic specifications \cite{fainekos2009robustness}. Apart from falsification, recent research has also focused on \emph{mining} of temporal properties for CPSs \cite{jin2015mining}, which can make it easier to understand what proper specifications could be, given simulations of a system. \red{A generalization of robustness is presented in a recent algebraic framework for runtime verification \cite{jakvsic2018algebraic}.}

\red{There exist several approaches to transform models between other design tools. In \cite{kekatos2018formal}, the author presents a way to go from informal requirements to hybrid models with the use of pattern templates. In \cite{balsini2017generation}, a method for generating Simulink monitors from formal requirements is presented -- this procedure is essentially the opposite of the one presented in this paper. In \cite{dragomir2018refinement}, a tool is introduced for translating Simulink models into theories in the proof assistant Isabelle \cite{nipkow2002isabelle}. To our knowledge, there has been no previous work transforming causal signal-based specifications into STL formulas, a transformation investigated in this paper.}


When it comes to improvements of the falsification process itself, previous work has defined a modified version of STL \cite{akazaki2015time}, and there has been discussion showing the need for similar modifications in industrial applications \cite{eddeland2017objective}. The main point has been to improve the robustness information from temporal operators by averaging the robustness inside the timed intervals in question. \red{Another novel approach, which can include different interpretations of robustness for temporal operators, views temporal logic as filtering \cite{rodionova2016temporal}. This connects the fields of temporal logic and signal processing and allows for new ways of analysis.  }

Several works \cite{dokhanchi2017vacuity} \cite{akazaki2016falsification} have designed methods for faster falsification of a specific sub-class of specifications, namely \emph{request-response specifications}. Recently, an extension to falsification has been proposed where meta-parameters of falsification, \emph{e.g.} the number of control points, are variable and put into an outer optimization problem \cite{aerts2018temporal}. 

Valued Booleans \cite{claessen2018using} is a recently-proposed logic that captures both the truth value of properties, as well as how severely the properties are falsified. In this paper, we use a version of Valued Booleans to enhance the capabilities of falsification. 

\subsection{Contributions}
The main contributions of this work are:

\begin{enumerate}[i)]
    \item transformation of causal signal-based requirements into STL specifications;
    \item application of Valued Boolean additive semantics to the falsification process;
    \item evaluation of additive semantics for falsification of benchmark requirements. 
\end{enumerate}

The rest of the paper is organized as follows: in Section \ref{sec:falsification}, STL and the falsification problem are defined. The latter is used to evaluate different robust semantics later on. In Section \ref{sec:signal_based}, we define a framework for translating causal signal-based specifications into STL. Section \ref{sec:vbools} details the logic of Valued Booleans, with two kinds of robust semantics. Section \ref{sec:experiments} compares the two robust semantics in falsification for a set of benchmark models, and in Section \ref{sec:conclusions} our conclusions are presented.

%% file: Sections/2_falsification.tex
\section{Signal Temporal Logic and Falsification}\label{sec:falsification}

The specification language STL is widely used for falsification of CPSs. We omit
the definition of the robust semantics of STL, as it is almost identical to the
\emph{max} semantics of VBools, which we define in Section
\ref{sec:max_semantics}. For details on STL, we refer the reader to other works
\cite{Donze2010}.

\subsection{Discrete-time signals}
Throughout this paper, we discuss specifications defined for signals and signal
values. The semantics of VBools is in terms of discrete-time signals, and for
the sake of consistency we also define STL this way, even though it is usually
defined in terms of continuous-time signals \cite{donze2010robust}. The main point of doing this is to make it clear how temporal operators can be defined in terms of conjunction, but generalizing to continuous time is possible \cite{fainekos2007robust} \cite{fainekos2009robustness}. \red{For practical purposes, falsification and monitoring of signals is performed on the output of simulated systems, where time has to be discretized to numerically solve the systems.}

\begin{mydef}
A \emph{discrete-time signal} is a function $x[k]$ from a finite subset of $I \subset \mathbb{Z}$ to $\mathbb{R}$, where $k \in I$. The set $I$ labels the time instants of the signals, and the signal takes on continuous values at each of those time instants. 
\end{mydef}

\subsection{Signal Temporal Logic}

The grammar of STL formulas is defined as 
\begin{align*}
\mu &:: = x < r \ | \ x \leq r \ | \ x \geq r \ | \ x > r \ | \ x = r\\
\varphi &:: = \mu \ | \  \lnot \mu \ | \  \varphi \land \psi \ | \ \Box_{[a,b]} \psi \ | \ \varphi \ \mathcal{U}_{[a,b]} \psi,
\end{align*}
where $\mu$ is a predicate, and $\varphi$ and $\psi$ are STL formulas. \red{$\land$ denotes logical \emph{and}, $\Box_{[a,b]}$ is the timed \emph{globally} (or \emph{always}) operator, and $\mathcal{U}_{[a,b]}$ is the timed \emph{until} operator. Due to De Morgan's laws, we can define logical \emph{or}} $\varphi \lor \psi$ as $\lnot(\lnot\varphi \land \lnot\psi)$. \red{There is a similar identity for the temporal operators, which lets us define timed \emph{eventually}} $\lozenge_{[a,b]}\varphi$ as $\lnot(\Box_{[a,b]}\lnot\varphi)$. \red{These identities will also be used in Section \ref{sec:vbools}}. Similarly to \cite{raman2015reactive}, we define the validity of a formula $\varphi$ with respect to the discrete-time signal $x$ at time instant $k$ as
\begin{align*}
    &(x,k) \models \mu &\Leftrightarrow \ \ &\mu(x[k])\\
    &(x,k) \models \neg\mu &\Leftrightarrow \ \ &\neg((x,k) \models \mu)\\
    &(x,k) \models \varphi \land \psi &\Leftrightarrow \ \ &(x,k) \models \varphi \land (x,k) \models \psi\\
    &(x,k) \models \varphi \lor \psi &\Leftrightarrow \ \ &(x,k) \models \varphi \lor (x,k) \models \psi\\
    &(x,k) \models \square_{[a,b]}\varphi &\Leftrightarrow \ \ &\forall k' \in [k+a,k+b], (x,k') \models \varphi\\
    &(x,k) \models \lozenge_{[a,b]}\varphi &\Leftrightarrow \ \ &\exists k' \in [k+a,k+b], (x,k') \models \varphi\\
    &(x,k) \models \varphi \ \mathcal{U}_{[a,b]}\psi &\Leftrightarrow \ \ &\exists k' \in [k+a,k+b] \ \ (x,k') \models \psi\\ \nonumber
    & & &\land \forall k'' \in [k,k'), (x,k'') \models \varphi
\end{align*}

We will provide an example of STL specification for clarity. The first example is a benchmark specification from \cite{bardh2014benchmarks}, informally stated as \textit{``During all simulation times, the engine speed $\omega$ and the vehicle speed $v$ never reach $\bar{\omega}$ and $\bar{v}$, respectively.''} The corresponding STL formula is 

\begin{equation*}
    \phi_2^{AT} = \Box((\omega < \bar{\omega}) \land (v < \bar{v})).
\end{equation*}

$\phi_2^{AT}$ contains two operators: $\Box$ and $\land$. The \emph{modal depth} of a formula is the deepest nesting of temporal operators (\emph{i.e.} $\Box, \lozenge, \mathcal{U}$) in it. For $\phi_2^{AT}$, the modal depth is 1.

\subsection{Falsification}
\begin{figure*}[!t]
\begin{center}
\input{figures/falsification.tex}
\caption{A flowchart describing a slightly modified version of the optimization-based falsification procedure of Breach. In this paper we deal with defining a specification transformer, as well as considering alternative robustness functions (the two shaded nodes).}
\label{fig:falsification}
\end{center}
\end{figure*}
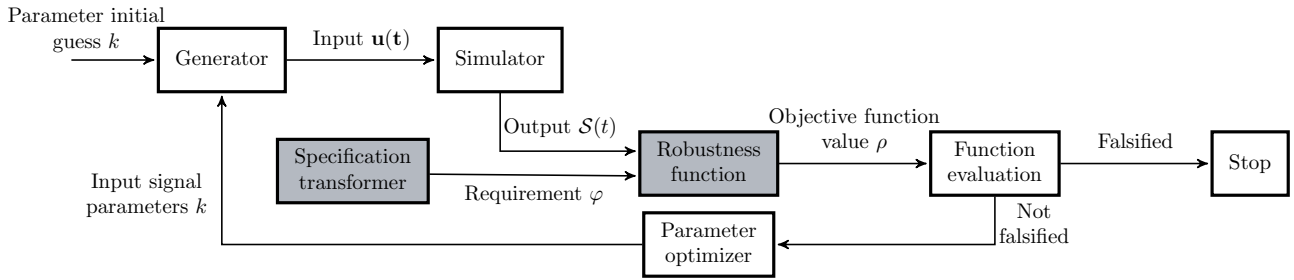

Temporal logic falsification is an approach to finding counterexamples to models of CPSs, given a specification in temporal logic. The problem of generating a test case for the CPS is treated as an optimization problem, where one attempts to minimize the robustness of the STL specification, given an input parametrization of the system. Figure \ref{fig:falsification} illustrates the main falsification procedure used in this paper (with the use of the tool Breach), which we have adapted to use VBools instead of STL robust semantics. 

The \emph{Generator} takes the input parametrization to generate an input to the system under test. The \emph{Simulator} generates a simulation trace, which is used together with the specification $\varphi$ to evaluate VBool robustness for the simulation. The VBool robustness is evaluated to see whether the specification is falsified or not. If it is not falsified, new parameters are sampled and the process is repeated. The \emph{Parameter Optimizer} is a global optimizer which attempts to find new input parameters that are closer to falsifying the specification, \emph{i.e.}, parameters that lead to a lower VBool robustness. 

In this work, we investigate two modifications to the falsification procedure. In Section \ref{sec:signal_based}, we introduce a transformation of signal-based requirements into STL (with a \emph{specification transformer}) as a means of allowing falsification to be performed by testers who are not used to temporal logic specifications. In Section \ref{sec:vbools}, the logic of VBools is introduced which allows the tester to control the objective function used in the falsification optimization problem.

%% file: figures/falsification.tex
\definecolor{green}{RGB}{40,188,0}
\definecolor{red}{RGB}{155,0,0}
\definecolor{darkgreen}{RGB}{8,137,0}
\definecolor{darkblue}{RGB}{9,52,226}
\definecolor{state3background}{RGB}{180, 185, 193}
\usetikzlibrary{arrows,automata,calc}
\tikzset{
    state/.style={
           rectangle,
           rounded corners,
           draw=black, very thick,
           minimum height=2em,
           inner sep=2pt,
           },
    state2/.style={
           rectangle,
           draw=black, very thick,
           minimum height=2em,
           inner sep=2pt,
           minimum size=3em
           },
    state3/.style={
           rectangle,
           draw=black, very thick,
           fill=state3background,
           minimum height=2em,
           inner sep=2pt,
           minimum size=3em
           },
}

\begin{tikzpicture}[->,>=stealth',shorten >=1pt,auto,semithick]

\node[state2] (1) [scale=0.8]{
    \begin{tabular}{c}
        Generator
    \end{tabular}};
    
\node[state2] (2) [right=2cm of 1,scale=0.8] {
    \begin{tabular}{c}
        Simulator
    \end{tabular}};
    
\node[state3] (3) [below right=0.5cm and 1cm of 2,scale=0.8] {
    \begin{tabular}{c}
        Robustness\\
        function
    \end{tabular}};

\node[state2] (4) [right=2cm of 3,scale=0.8] {
    \begin{tabular}{c}
        Function\\
        evaluation
    \end{tabular}};
    
\node[state2] (5) [right=2cm of 4,scale=0.8] {
    \begin{tabular}{c}
        Stop
    \end{tabular}};

\node[state2] (6) [below=0.2cm of 3,scale=0.8] {
    \begin{tabular}{c}
        Parameter\\
        optimizer
    \end{tabular}};
   
\node[state3] (7) [below left=0.6cm and 0.1cm of 2,scale=0.8] {
    \begin{tabular}{c}
         Specification\\
         transformer
    \end{tabular}};


    
\node at (4.5,-0.9) [scale=0.8]{
        \begin{tabular}{c}
            Output $\mathcal{S}(t)$
        \end{tabular}};
        
\node at (10.8,-2.2) [scale=0.8]{
        \begin{tabular}{c}
            Not\\
            falsified
        \end{tabular}};
        
\node at (-1,-1.8) [scale=0.8]{
        \begin{tabular}{c}
            Input signal\\
            parameters $k$
        \end{tabular}};
        
\node at (-1.8,0.4) [scale=0.8]{
        \begin{tabular}{c}
            Parameter initial\\
            guess $k$
        \end{tabular}};
    
\coordinate[below left=1.08cm and 0.1cm of 2] (d1);

\path 
    (1) edge []             node[scale=0.8] {
        \begin{tabular}{c}
            Input $\mathbf{u(t)}$
        \end{tabular}} (2)
    (3) edge []             node[scale=0.8] {
        \begin{tabular}{c}
            Objective function\\
            value $\rho$
        \end{tabular}} (4)
    (d1) edge [below]             node[scale=0.8] {
        \begin{tabular}{c}
            Requirement $\varphi$
        \end{tabular}} (3.190)
    (4) edge []             node[scale=0.8] {
        \begin{tabular}{c}
            Falsified
        \end{tabular}} (5);
    
    
    \draw[->] (2.south) |- (3.170);
    \draw[->] (4.south) |- (6.east);
    \draw[->] (6.west) -| (1.south);
    \draw[->] (-2,0) -- (1.west);
    
\end{tikzpicture}

%% file: Sections/3_signal_based_specifications.tex
\section{Signal-Based Specifications}\label{sec:signal_based}
As has been noted before \cite{dokhanchi2015metric}, writing specifications in temporal logic is not trivial. Approaches that have been used to solve this problem are creating tools that make it easier to write specifications \cite{hoxha2015vispec}, automatically detecting faulty specifications \cite{dokhanchi2018formal}, and defining template specifications to make it easier for testers to formulate their requirements formally \cite{kapinski2016st}. In this paper, we instead allow test engineers to write specifications in a formalism they already know, namely a causal signal-based framework (in our case using Simulink \cite{simulink}). 

The main idea behind a signal-based safety specification is to directly take signals from the simulated system, then using different operators (blocks) to give an output signal that \emph{at each simulated time instant} is either 1 (specification is fulfilled) or 0 (specification is not fulfilled). The advantage of this is that a test can easily be automatically executed and evaluated at the same time as the system itself is simulated. 

By using a signal-based specification, we exploit the fact that the test engineers are experts at expressing specifications in, for example, Simulink.
The drawback is that a signal-based specification does not compute robustness
values, and so can not be directly used for falsification. To solve this
problem, we automatically translate signal-based specifications into STL
formulas to be used by Breach.

\subsection{STL specifications in a signal-based framework}

As an example, we wish to show an implementation of a version of $\phi_1^{AT}$ from \cite{bardh2014benchmarks}, which is defined as

\begin{equation}
    \phi_1^{AT} = \Box(\omega < \bar{\omega}).
\end{equation}

\begin{figure}[!t]
\centering
\includegraphics[width=2.5in]{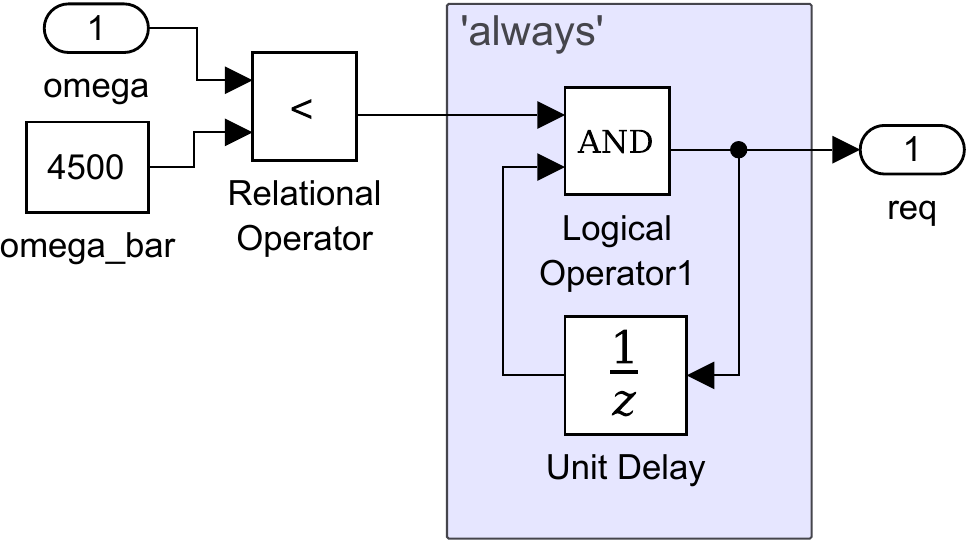}
\caption{A simple example of a specification expressed in Simulink. The natural language interpretation is \textit{``During all simulation times $t \in [0,T]$, the engine speed $\omega$ never reaches $\bar{\omega}$''}. For the implementation to be correct, the initial condition of the \emph{Unit Delay} block must be non-zero. }
\label{fig:example1}
\end{figure}

It should be noted that since a specification implemented in Simulink must be causal, temporal operators that look forward in time cannot be explicitly modeled. However, a specification model with a similar meaning to $\phi_1^{AT}$ (with $\bar{\omega} = 4500$) is presented in Figure \ref{fig:example1}. 

Assume that the specification is evaluated on a simulation trace with a finite set of sampled data points $K$. The interpretation of the signal \texttt{req} at sample $k \in K$ is then

\begin{equation}
    \texttt{req}[k] = 
    \begin{cases}
        \top & \text{if  } \omega[k'] < 4500,  \forall k' \in K \cap [0,k]\\
        \bot & \text{otherwise,}
    \end{cases}
\end{equation}

while the Boolean evaluation of the STL formula $\phi_1^{AT}$ at time $k$ can be informally expressed as

\begin{equation}
    \phi_1^{AT}[k] = 
    \begin{cases}
        \top & \text{if  } \omega[k'] < 4500, \forall k' \in K \cap [k, k+T]\\
        \bot & \text{otherwise.}
    \end{cases}
\end{equation}

As can be easily seen, $\texttt{req}(k)$ is not equal to the Boolean evaluation of $\phi_1^{AT}(k)$ for all $k$, but $\texttt{req}(T) = \phi_1^{AT}(0)$. This is the only thing that is needed to achieve equivalence between the Boolean interpretation of a causal signal-based requirement and its STL equivalent, since the STL formula will be evaluated for time 0, and the signal-based specification will be evaluated at the final simulation time. \red{We note that another possible approach to generate specifications in this setting would be to consider past-time operators of STL, instead of future-time operators as presented here. }

\subsection{Signal-based specifications expressed in STL}

The goal is to be able to take any signal-based specification, and then transform it into an STL formula so that it can be used for falsification. Ideally, each signal in the signal-based model would be assigned an STL formula, but since the semantics of a signal-based framework are not typically equivalent to the semantics of STL, they have different levels of expressivity.

In this section, a \emph{Signal} is a variable that has defined values over time, and it can be a scalar or a vector. A Signal corresponds to a signal in a causal model. A \emph{Formula} is a special case of a Signal, namely a Signal that always has a Boolean value (\emph{i.e.} it is either true or false).

To model signals whose behaviour varies depending on the value of a Boolean expression, we define the types \emph{FormulaTable} and \emph{SignalTable} as
\begin{align}
    FormulaTable &= \mathcal{P}(Formula \times Formula)\\
    SignalTable &= \mathcal{P}(Formula \times Signal),
\end{align}

where $\mathcal{P}$ denotes the powerset operation. A FormulaTable or
SignalTable consists of a set of entries, where each entry is a pair of a
precondition, expressed as an STL formula, and a consequent, which is the value
taken by the formula or signal when the precondition is true. The disjunction of
all preconditions for any FormulaTable or SignalTable must be $\top$.\footnote{In
particular, if a FormulaTable or SignalTable only has one entry, the
precondition in that entry must be $\top$.}

\begin{figure}[!t]
\centering
\includegraphics[width=2.5in]{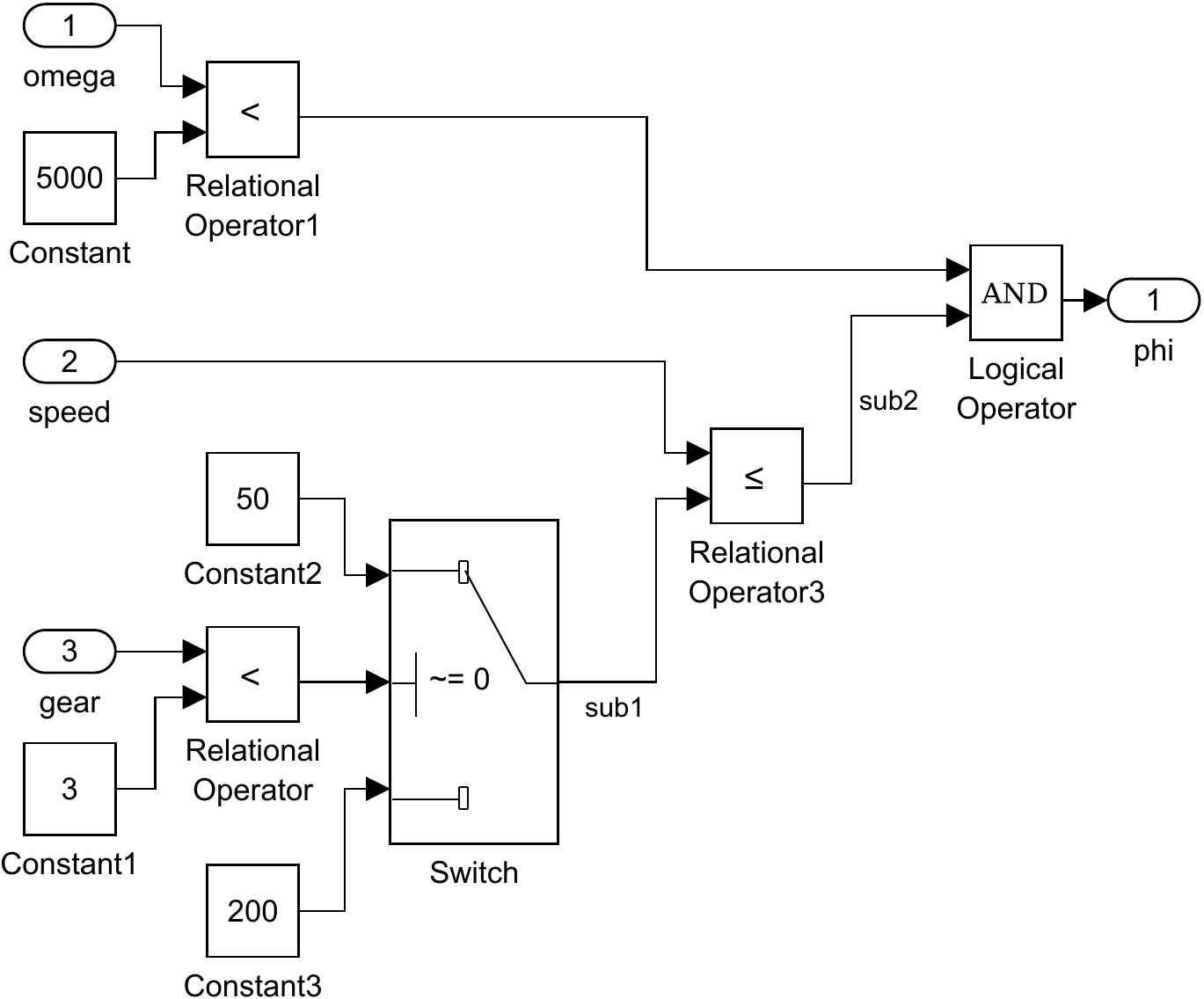}
\caption{An example of a requirement with a conditional statement, implemented with the use of a Simulink Switch block. }
\label{fig:example3}
\end{figure}

Figure \ref{fig:example3} shows a Simulink encoding of the natural language
requirement \emph{``The engine speed $\omega$ should always be below 5000 RPM.
Additionally, if we are in third gear or lower, the speed $v$ should be below 50
km/h; otherwise, the speed should be below 200 km/h.''} The Switch block assigns
a value to its output signal according to the rule:
\begin{algorithmic}
    \If {$gear \leq 3$} 
        \State $sub1 = 50$
    \Else
        \State $sub1 = 200$
    \EndIf 
\end{algorithmic}

The signal $sub1$ is translated into a SignalTable, shown in Table
\ref{tab:sub1}. The signals $sub2$ and $phi$ are translated into FormulaTables,
seen in Tables \ref{tab:sub2} and \ref{tab:finalSTL} respectively. Since there
are two conditions, the SignalTables and FormulaTables have two entries. The
SignalTable for $sub1$ has two entries because it is the output of a Switch
block; the FormulaTables for $sub2$ and $phi$ have two entries because the
FormulaTable for the output of a block has an entry for each possible
combination of preconditions from the block's inputs.

\begin{table}[!t]
\renewcommand{\arraystretch}{1.3}
\caption{The SignalTable for the signal $sub1$ in the Example in Figure \ref{fig:example3}.}
\label{tab:sub1}
\centering
\begin{tabular}{|c|c|}
\hline
Precondition & Consequent\\
\hline
\hline
$gear < 3$ & $50$ \\
\hline
$\lnot(gear < 3)$ & $200$ \\
\hline
\end{tabular}
\end{table}

\begin{table}[!t]
\renewcommand{\arraystretch}{1.3}
\caption{The FormulaTable for the signal $sub2$ in the Example in Figure \ref{fig:example3}.}
\label{tab:sub2}
\centering
\begin{tabular}{|c|c|}
\hline
Precondition & Consequent\\
\hline
\hline
$gear < 3$ & $v < 50$ \\
\hline
$\lnot(gear < 3)$ & $v < 200$ \\
\hline
\end{tabular}
\end{table}

\begin{table}[!t]
\renewcommand{\arraystretch}{1.3}
\caption{The FormulaTable for the output $phi$ in the Example in Figure \ref{fig:example3}.}
\label{tab:finalSTL}
\centering
\begin{tabular}{|c|c|}
\hline
Precondition & Consequent\\
\hline
\hline
$gear < 3$ & $(\omega < 5000) \land (v < 50)$ \\
\hline
$\lnot(gear < 3)$ & $(\omega < 5000) \land (v < 200)$ \\
\hline
\end{tabular}
\end{table}


\red{To transform a binary operator}\footnote{A unary operator is a simplification of the
algorithm presented. An $n$-ary operator, for example $\land$, is implemented
pairwise (meaning that $a \land b \land c$ is transformed to $(a \land b) \land
c$, which is possible due to associativity of both max and additive semantics).}\red{, we construct the following table:}
\begin{multline*}
  \{ (prereq1 \land prereq2, operator(conseq1,
conseq2)) \\ \mid (prereq1, conseq1) \in in1, (prereq2, conseq2) \in in2
\}.
\end{multline*}
\red{As can be seen, the operator of the block is applied to each consequent of the table.} The number of entries $\alpha$ in the table that is produced from a
block with $K$ inputs $u_1, u_2, \ldots, u_K$ will be $\Pi_{k=1}^K
\alpha_{u_k}$, where $\alpha_{u_k}$ is the number of entries in the table of
input $u_k$.



An important difference between signal-based specifications and STL specifications is due to conditional blocks. The archetypical conditional block is the \emph{Switch} block, which takes three inputs and lets the output be either the first or the third input, depending on a user-defined condition on the second input. The output table of a Switch block has $\alpha = \alpha_2(\alpha_1 + \alpha_3)$ entries.


To translate a FormulaTable into an STL formula, one can consider the ``STL semantics'' for a Simulink switch (with inputs $x_1, x_2, x_3$) as either
\begin{equation}
    (x_2 \land x_1) \lor (\lnot(x_2) \land x_3)
\end{equation}
or
\begin{equation}
    (x_2 \implies x_1) \land (\lnot(x_2) \implies x_3).
\end{equation}

Note that these two expressions are logically equivalent, but they do not necessarily yield the same robustness value.

\subsection{Recursive loops in specifications}

To transform a signal-based specification into STL, we perform a backwards depth-first search from the output of the specification, assigning a FormulaTable or SignalTable to each signal in the specification. For simple specifications \red{without loops}, the search algorithm discussed will terminate and assign an STL formula to the signal leading to the outport of the specification. However, any kind of temporal behaviour in a specification is typically implemented as a recursive loop, which leads to the basic search algorithm not terminating -- something that needs to be taken care of when transforming the STL formula. 

\subsubsection{Handling recursive loops, approach 1}

If the length of the simulation is known and finite, we can transform a
recursive loop into a formula that explicitly computes its value in terms of the
values at all earlier time steps. For the example presented in Figure
\ref{fig:example1}, this corresponds to the final output

\begin{equation}\label{eq:large_req}
    req(k) = \bigwedge_{k' = 0}^{k}(\omega(k') < \bar{\omega}).
\end{equation}

However, this results in large and potentially unreadable STL formulas as soon as there is some recursion involved, even for simple specifications. For example, given a simulation time in $[0,10]$ and a fixed simulation step time of $0.01$, requirement \eqref{eq:large_req} results in an STL specification with $1001$ $\land$-connectives, and more than $31 000$ characters when written in Breach syntax. Even though the robustness values for the formula will still be the same as for the STL formula $\varphi = \Box_{[0, 10]}(\omega(t) < \bar{\omega})$, we typically want something that is as readable as possible.

\subsubsection{Handling recursive loops, approach 2}
If it is a goal to keep the automatically transformed STL formulas as short as possible, we use \emph{templates} of combinations of different temporal operators that are implemented as their own subsystems in the model. This is in a way very similar to ST-Lib \cite{kapinski2016st}, but instead of defining templates that can be used to build specifications from the ground up directly in STL, we define templates in Simulink that are associated to predefined STL formulas. 

For the example in Figure \ref{fig:example1}, one such template could be the $\Box$ operator, which in practice would be a subsystem replacing the blocks in the shaded area.

\subsubsection{Handling recursive loops, approach 3}\label{sec:recursive_practice}
A final possibility is to treat a recursive loop as a black box rather than
translating it to STL. To do this, we treat the output of the delay
block\footnote{Note that a delay block must be present in the loop, otherwise it
would be an algebraic loop.} as a signal in the specification, \emph{i.e.}
consider anything before the delay block to be part of the model. The value of
the signal is computed by the model, and the STL specification simply refers to
the signal. This approach is useful when we want to avoid the inefficient
encoding of approach 1 and the recursive loop does not correspond to a
predefined template. It is also needed when a block applies a general function
to its input, in which case the function output cannot be explicitly defined as a formula, but
by treating the function as part of the system we are still able to translate the specification to STL.

\red{To summarize our implementation, whenever there is a recursive loop, approach 3 will be used unless there is a template defined for the part of the model containing the loop. If there is a template, approach 2 is used. This means that the specification transformation is fully automatic, with the possibility to include more detailed information about the specification by the use of templates. }

\begin{figure}[!t]
\centering
\includegraphics[width=2.5in]{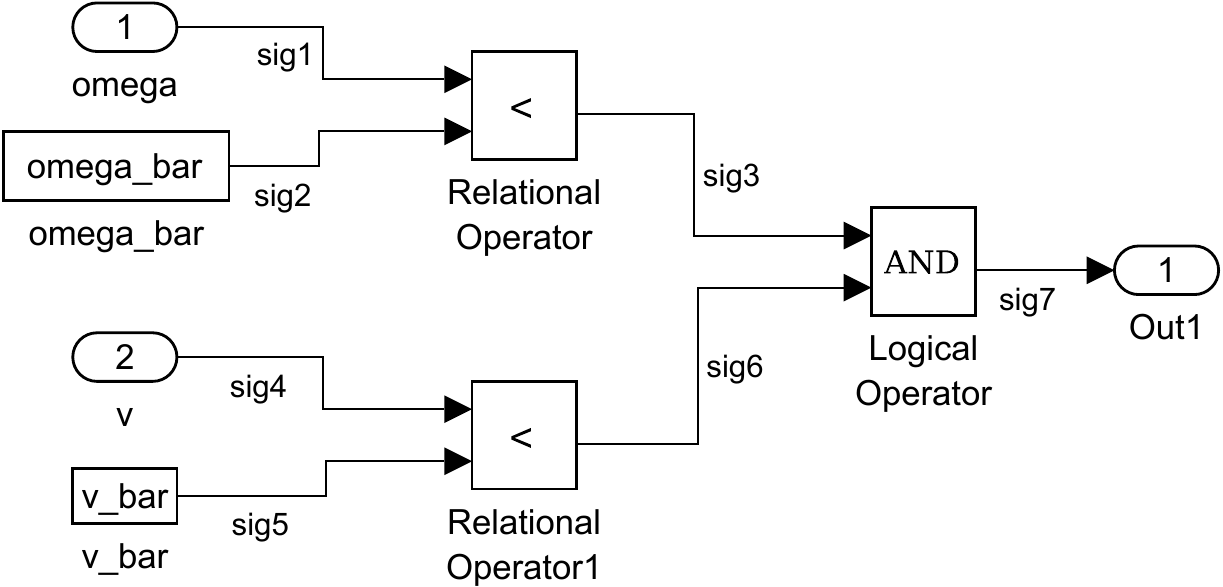}
\caption{An implementation of the STL specification $(\omega < \bar{\omega}) \land (v < \bar{v})$, which is interpreted as \textit{"The engine speed $\omega$ and the vehicle speed $v$ never reach $\bar{\omega}$ and $\bar{v}$, respectively"}.}
\label{fig:example2}
\end{figure}

An extended example of this can be shown by considering the signal-based specification in Figure \ref{fig:example2}. The specification itself is part of $\phi_{2}^{AT}$ \cite{bardh2014benchmarks}. Some different ways to interpret this specification, based on which of the given signals are considered as part of the model (\emph{logged signals}), are shown in Table \ref{tab:spec_interpretations}. 

\begin{table}[!t]
\renewcommand{\arraystretch}{1.3}
\caption{Some Possible Interpretations of Specification in Figure \ref{fig:example2}.}
\label{tab:spec_interpretations}
\centering
\begin{tabular}{|c|c|}
\hline
Logged signals & STL Formula\\
\hline
\hline
- & $(\omega < \bar{\omega}) \land (v < \bar{v})$ \\
\hline
sig1 & $(sig1 < \bar{\omega}) \land (v < \bar{v})$ \\
\hline
sig2 & $(\omega < sig2) \land (v < \bar{v})$ \\
\hline
sig1, sig4 & $(sig1 < \bar{\omega}) \land (sig4 < \bar{v})$ \\
\hline
sig3 & $\lnot(sig3 = 0) \land (v < \bar{v})$ \\
\hline
sig6 & $(\omega < \bar{\omega}) \land \lnot (sig6 = 0)$ \\
\hline
sig3, sig6 & $\lnot(sig3 = 0) \land \lnot(sig6 = 0)$ \\
\hline
sig7 & $\lnot(sig7 = 0)$ \\
\hline
\end{tabular}
\end{table}

The advantage of this approach is that we can be certain to translate \emph{any} signal-based specification to STL, while the disadvantage is that the generated STL specification might be less suited to falsification than had
we translated the recursive loops to STL. For example, the specification $(\omega < \bar{\omega}) \land (v < \bar{v})$ (for the Automatic Transmission benchmark) has many possible robustness values since the signals $\omega$ and $v$ have many different potential values. However, the specification $\lnot(sig7 = 0)$ (which has exactly the same Boolean truth value) only has two possible robustness values. This makes falsification harder, since the optimization solver will not see how close the specification came to failing.

\red{The claim above about being able to translate any signal-based specification to STL is a very strong one, as there are specifications that can be modeled using e.g. Simulink that cannot be expressed in STL, see \cite{brim2014stl}. However, the solution presented here is that if there is a block that is not expressible in STL, its output will be logged (and the block is therefore not explicitly stated using STL, as that would be impossible). Also, specific temporal behaviours that are not expressible in STL result in logging of certain blocks as explained earlier in this section, which means that we can indeed transform any requirement to STL, however parts of the requirement may not be stated explicitly. Providing a formal proof of the correctness of the translation is beyond the scope of this paper, and instead considered future work. We note, however, that such a formal proof may be problematic due to non-standard semantics of Simulink \cite{benveniste2012non}.}

\subsection{When semantics do not match}
For the specification transformation framework presented in this paper, there is a difference between logical formulas and signals. However, in a signal-based setting there is not, so it is possible for a block to get the wrong type of input. For example, consider the expected inputs and outputs of the following blocks:
\begin{align*}
    \land &: FormulaTable \times FormulaTable \to FormulaTable\\
    < &: SignalTable \times SignalTable \to FormulaTable\\
    + &: SignalTable \times SignalTable \to SignalTable
\end{align*}

There are two cases for unexpected input types: either a SignalTable is provided when a FormulaTable should be, or a FormulaTable is provided when a SignalTable should be. 

\subsubsection{SignalTable provided instead of FormulaTable}
This can occur if, for example, we apply the $\land$ operator to two real-valued signals $x$ and $y$. Simulink (and MATLAB) semantics interpret the Boolean evaluation of these signals as being false if they are equal to zero, and true otherwise. This means that we can transform a SignalTable to a FormulaTable by comparing equality of the SignalTable's consequent to zero, and then applying the $\lnot$ operator. This is accomplished by the $S2F$ function:
\begin{align*}
    S2F &: SignalTable \to FormulaTable\\
    S2F &(\langle precond, conseq\rangle)= \langle precond, \lnot(conseq = 0)\rangle
\end{align*}

\subsubsection{FormulaTable provided instead of SignalTable}
This can occur if, for example, we try to add (using the $+$ operator) two predicates, such as $x>0$ and $y<10$. The meaning of this is clear when interpreted as signals according to the Simulink semantics: the output of the $+$ operator will have value 0 (when both predicates are false), 1 (when exactly one of the predicates are true), or 2 (when both predicates are true). However, in STL we cannot define a formula by adding logical formulas together. 

In this case, if the sum is later used as a formula by comparing it to zero
(\emph{i.e.} the signal expression to be evaluated is $\big((x>0) + (y<10)\big)
= 0$), then an equivalent STL formula would be $\lnot((x>0) \lor (y<10))$. However, it is not clear how to generalize this observation, so instead we consider anything before the block in question (here, the $+$ operator) to be a  black box, and the output of the block is treated as a signal, using the same method described in Section \ref{sec:recursive_practice}.

%% file: Sections/4_vbools.tex
\section{Valued Booleans}\label{sec:vbools}

Valued Booleans (VBools) \cite{claessen2018using} is a logical framework in
which the tester can customize how robustness is computed by choosing between
several possible semantics for each connective. The semantics that are currently available are
a \emph{max} semantics (which is essentially the same as STL) and an \emph{additive} semantics.

A VBool is formally defined as a pair of a Boolean value and a robustness value. The robustness is a non-negative number, which may be infinite:
\begin{equation*}
    \mathbb{V} = \mathbb{B} \times \mathbb{R}_{\geq 0}
\end{equation*}

Note the difference between VBools and STL. In STL, there is no explicit Boolean value, but the robustness may be negative, and negative robustness represents falsehood. For VBools, the Boolean value is explicit and robustness may not be negative.

The VBool comparison operator $\leq_v$ is defined as:
\begin{align*}
\leq_v &: \mathbb{R} \times \mathbb{R} \rightarrow \mathbb{V} \\
x \leq_v y &=
  \begin{cases}
    (\top, y-x) & \mathrm{ if } \ \ x \leq y \\
    (\bot, x-y) & \mathrm{ otherwise.}
  \end{cases}
\end{align*}

$\top$ and $\bot$ denote true and false, respectively. The other comparison operators are defined in terms of $\leq_v$, except for $=_v$ which is defined as
\begin{equation*}
    x =_v y =
    \begin{cases}
        (\top, K) & \mathrm{ if } \ \ x = y\\
        (\bot, K) & \mathrm{ otherwise,}
    \end{cases}
\end{equation*}

where $K$ is an arbitrary constant. Truth values and negation are defined as
\begin{align*}
\top_{\!\mathrm{v}} &= (\top, \infty) \\
\bot_{\mathrm{v}} &= (\bot, \infty) \\
\lnot_v (b, x) &= (\lnot b, x).
\end{align*}

The rest of the operators are defined in two different ways. One is called \emph{max}
semantics and the other \emph{additive} semantics.

\subsection{Max semantics}\label{sec:max_semantics}
The \emph{max and} operator is defined as

\begin{align*}
    (\top, x) \land_{max} (\top, y) &= (\top, min(x,y)) \\
    (\bot, x) \land_{max} (\top, y) &= (\bot, x) \\
    (\top, x) \land_{max} (\bot, y) &= (\bot, y) \\
    (\bot, x) \land_{max} (\bot, y) &= (\bot, max(x,y)).
\end{align*}

The first clause models the idea that in order to falsify $x \land y$, it is
enough to falsify whichever of $x$ and $y$ has the lowest robustness. If we are in the
second clause, then $x \land y$ is false, and in order to make it true, we must make $x$ true; the
third clause is similar. The final clause is dual to the first clause: in order
to make $x \land y$ true we must make both $x$ and $y$ true, and the robustness is determined by whichever of $x$ and $y$ seems to be hardest to make true, i.e., has the highest robustness as a false VBool.

The \emph{max or} operator is defined in terms of the \emph{max and} operator: $(b_x, x) \lor_{max} (b_y, y) = \lnot_v ( \lnot_v (b_x, x) \land_{max} \lnot_v (b_y, y))$.

The timed \emph{max always} operator (over the interval $[a,b]$) is also defined in terms of the \emph{max and} operator as

\begin{align*}
    \Box_{max,[a,b]}\varphi = \sideset{}{_{max}}\bigwedge\limits_{k = a}^b \varphi[k],
\end{align*}

where $\varphi$ is a finite sequence of VBools defined for all the discrete time instants in $[a,b]$.

The timed \emph{max eventually}-operator is defined as $\lozenge_{max,[a,b]}\varphi = \lnot(\Box_{max,[a,b]}(\lnot_v\varphi))$.
Finally, for completeness we also define the \emph{max until}-operator as

\begin{align*}
    \varphi \ &\mathcal{U}_{max,[a,b]} \ \psi \\
    & = \sideset{}{_{max}}\bigvee \limits_{k=a}^{b}\left(\psi \land_{max} \left(\sideset{}{_{max}}\bigwedge\limits_{k'=a}^{b-1}\varphi[k']\right)\right).
\end{align*}

It can be seen that the max semantics for VBool are almost equivalent to the robust semantics of STL, with the only difference being that VBools distinguish between ``true with robustness 0'' and ``false with robustness 0'', while STL does not. \red{This difference is only technical and in practice the two semantics behave the same. However, a single VBool formula can contain both max connectives and connectives using other semantics, such as the additive semantics defined below.} 

\subsection{Additive semantics}

The \emph{additive and}-operator is defined as\footnote{\red{In the case where both x and y are true, but either $x$ or $y$ is 0, we define the resulting robustness to be 0. This to avoid division by 0, and 0 is also the limit of the expression as $x$ or $y$ goes to 0.}}

\begin{align*}
    (\top, x) \land_{+} (\top, y) &= \left(\top, \frac{1}{\frac{1}{x} + \frac{1}{y}}\right) \\
    (\bot, x) \land_{+} (\top, y) &= (\bot, x) \\
    (\top, x) \land_{+} (\bot, y) &= (\bot, y)\\
    (\bot, x) \land_{+} (\bot, y) &= (\bot, x+y).
\end{align*}

As with the max semantics, the additive semantics for $\land$ is based on the observation that in order to falsify $x \land y$, it is enough to falsify either $x$ or $y$. The first clause is inspired by the formula for parallel resistance; the formula $1/(1/x+1/y)$ gives a robustness which is less than the maximum of $x$ and $y$. It roughly models the idea that although we need only falsify one of $x$ and $y$, we do not know which one of them can be falsified. The second and third clauses are the same as in the max semantics. By using addition in the fourth clause rather than $max$, we model the idea that in order to make $x \land y$ true, we need to make both $x$ and $y$ true, not just whichever of them has the highest robustness.

The \emph{additive or}-operator is defined as $(b_x, x) \lor_+ (b_y, y) = \lnot_v ( \lnot_v (b_x, x) \land_+ \lnot_v (b_y, y))$, and the timed \emph{additive always}-operator (over the time interval $[a,b]$) is defined (similar to the $max$ case) as
\begin{align*}
    \Box_{+,[a,b]}\varphi = \sideset{}{_{+}}\bigwedge\limits_{k = a}^b (\varphi[k] \#' \delta t),
\end{align*}
where $\varphi$ is a finite sequence of VBools defined for the time instants in $[a,b]$, $\delta t$ is the simulation step time for the time point in question, and $\#'$ is defined as
\begin{align*}
    (\bot, x) \#' k &= (\bot, x \cdot k)\\
    (\top, x) \#' k &= (\top, x/k).
\end{align*}
The use of $\#'$ makes the robustness independent of the simulation time step,
and means that the robustness of $\Box_{+,[a,b]} \varphi$, if $\varphi$ is false over the interval $[a,b]$, is equal to the \emph{integral} of the robustness of $\varphi$ over $[a,b]$.

The timed \emph{additive eventually}-operator is defined as $\lozenge_{+,[a,b]}\varphi = \lnot(\Box_{+,[a,b]}(\lnot_v\varphi))$ .

The \emph{additive until}-operator is defined as
\begin{align*}
    \varphi \ &\mathcal{U}_{+,[a,b]} \ \psi \\
    &= \sideset{}{_{+}}\bigvee \limits_{k=a}^{b}\left(\left(\psi[k]\#'\delta t\right) \land_{+} \left(\sideset{}{_{+}}\bigwedge\limits_{k'=a}^{b-1}\left(\varphi[k']\#'\delta t\right)\right)\right).
\end{align*}

Implication is defined slightly differently than in classical logic:
\[
  \phi \rightarrow_+ \psi = \lnot (\phi \# k) \lor \psi.
\]
Here $k$ is an arbitrary constant, and $\#$ scales the robustness of its argument:
\begin{align*}
    (\bot, x) \# k &= (\bot, x \cdot k)\\
    (\top, x) \# k &= (\top, x \cdot k).
\end{align*}
By scaling the left-hand side of the implication, we encourage the parameter optimizer to make the left-hand side true before trying to falsify the right-hand side.

\subsection{Properties for reasoning about Valued Booleans}

Most Valued Boolean connectives have two possible semantics, and the tester must
choose one of the semantics for each connective in the specification. The max
semantics corresponds closely to the existing robust semantics of STL \red{(and for use in falsification, they yield the same result)}, but the
additive semantics is entirely different. \red{The purpose of using max semantics for Valued Booleans instead of STL robustness is so that the tester can freely change between different robust semantics of VBools, even within a single formula}. This section compares the two
semantics of Valued Booleans and describes the different properties they have
which explain why a tester might choose to use one or the other. \red{For a more thorough discussion on Valued Booleans, we refer the reader to the work which introduced them \cite{claessen2018using}. This section rather discusses the practical issues of using Valued Booleans in the case of falsification.}

The ultimate goal of a robust semantics is to guide the falsification in the
right direction. Therefore, when a change in the input to the system brings a formula closer to being falsified, the robustness of the formula should go down.
This is the property we ideally want from a robust semantics. It is not always achievable in reality (because we can never be sure if we are really moving closer to a counterexample or not), but the more often it holds, the better. We are particularly interested in two special cases of this property, \emph{monotonicity} and \emph{sensitivity}. \red{For simplicity we assume that formulas are in negation normal form, i.e., negation only occurs as part of an atomic formula.}

\begin{mydef}
A formula is \textit{monotonic} if, when the robustness of some atomic subformula decreases (leaving the others unchanged), the robustness of the formula does \emph{not} increase.  
\end{mydef}

A nonmonotonic formula is disastrous for falsification as the parameter
optimizer will, moving from a test case to a strictly better one, observe the better test case as being worse instead. All VBool formulas are monotonic.

\begin{mydef}
A formula is \emph{sensitive} if changing the robustness of some atomic subformula (leaving the others unchanged) causes a change in the robustness of the formula. This captures the idea that if the output of the system changes then the robustness of the formula should usually change.
\end{mydef}

\subsubsection{\red{Importance of sensitivity for falsification}}

Sensitivity is vital for falsification because measuring changes in robustness
is how the parameter optimizer explores the input space. For falsification it
is only important that \emph{true} formulas be sensitive \red{if the falsification stops immediately when the counter-example is found (and robustness thus is equal to 0).} If moving from a test case to a strictly better test case does not affect robustness, then the parameter optimizer will not know when it has found a better test case.
The traditional semantics of Boolean logic is completely insensitive, which is
why a robust semantics is needed for falsification.

Unfortunately, the max semantics is \emph{not} sensitive: only one parameter of $p \land q$ is taken into account for any given test case. For example, if $p \land q$ is true, then the semantics is only sensitive to changes in whichever of $p$ and $q$ has the \emph{lowest} robustness.

The additive semantics is sensitive for many true formulas. For example, $p \land q$ is fully sensitive when $p$ and $q$ are both true. This means that, when falsifying the conjunction of several formulas, the parameter optimizer is able to observe changes in the robustness of any of the subformulas.
However, the formula $p \lor q$ is not fully sensitive when exactly one of its arguments is true.

\begin{figure*}[!t]
\begin{center}
  \subfloat[An example of a signal defining a property $\phi$. Robustness is lowest at about 48 seconds.]{\includegraphics[width=0.45\textwidth]{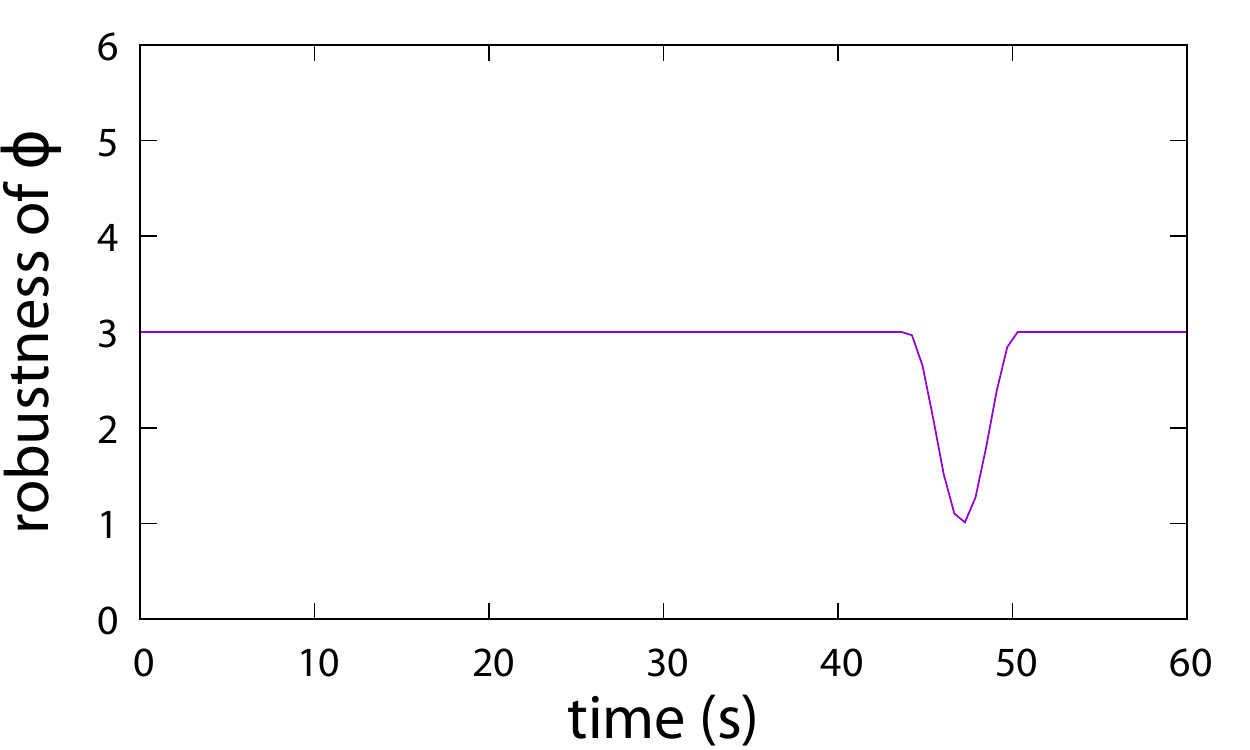}}
  \hskip0.05\textwidth
  \subfloat[In this trace, the property comes close to being falsified twice, at 20 seconds and 48 seconds. The ``max'' semantics assigns this trace the same robustness as for figure (a), even though it is strictly worse. The ``+'' semantics assigns it a lower robustness.]{\includegraphics[width=0.45\textwidth]{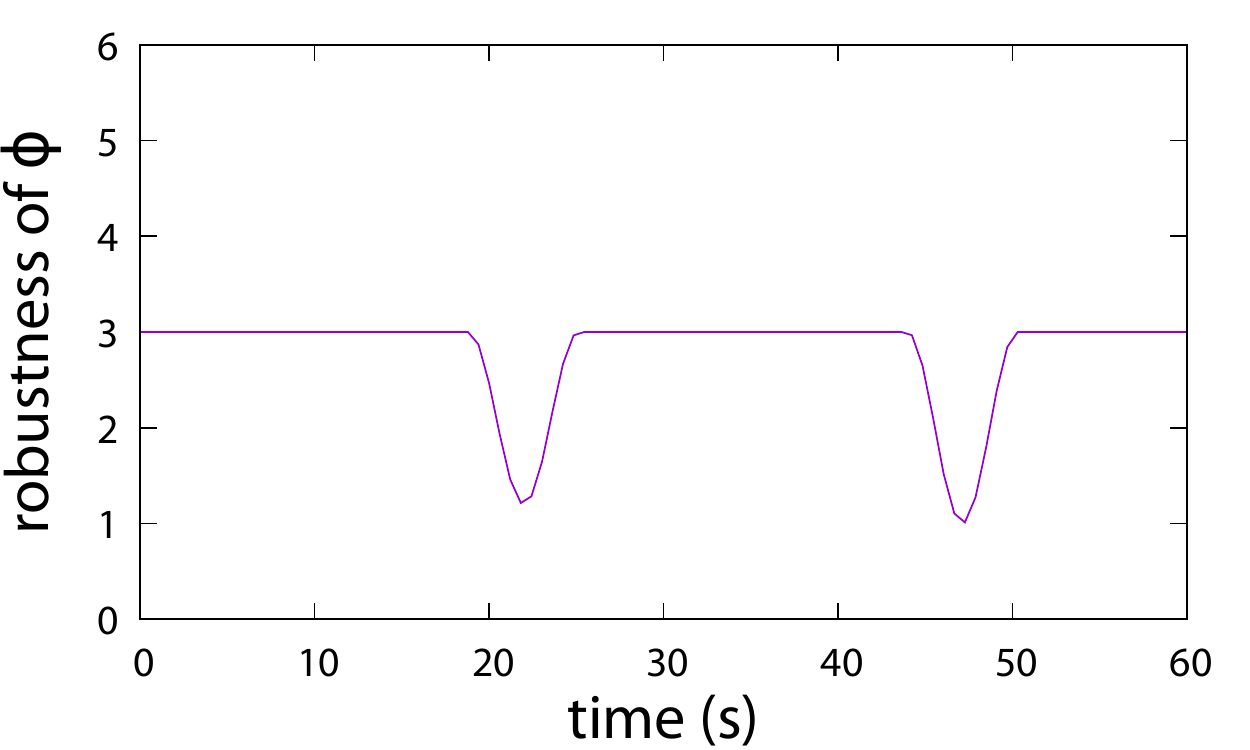}}

  \subfloat[This trace is similar to figure (a), but the robustness in the initial part of the trace is even higher. The ``+'' semantics assigns this trace a higher robustness than figure (a), even though it appears much closer to being falsified.]{\includegraphics[width=0.45\textwidth]{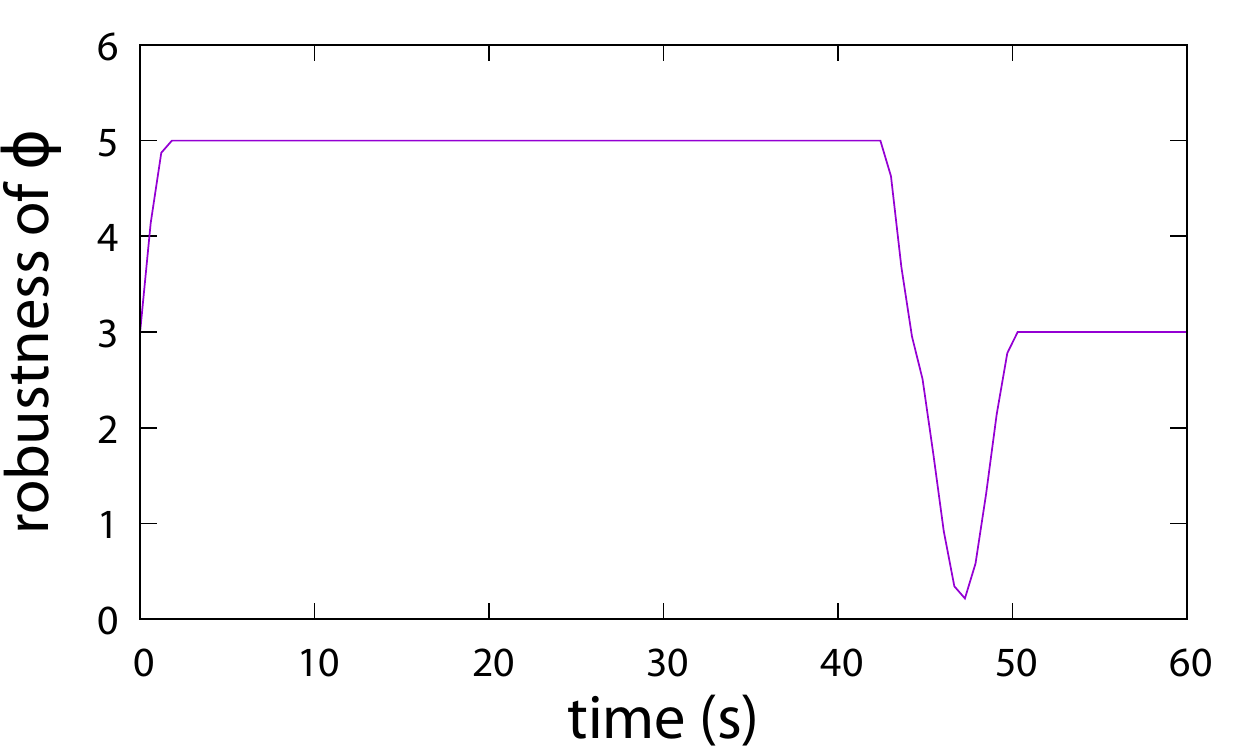}}
  \hskip0.05\textwidth
  \subfloat[In this trace, the property comes very close to being falsified at 48 seconds, but is more robustly true the rest of the time. The ``max'' semantics assigns this trace a lower robustness than figure (a). The ``+'' semantics also assigns it a slightly lower robustness.]{\includegraphics[width=0.45\textwidth]{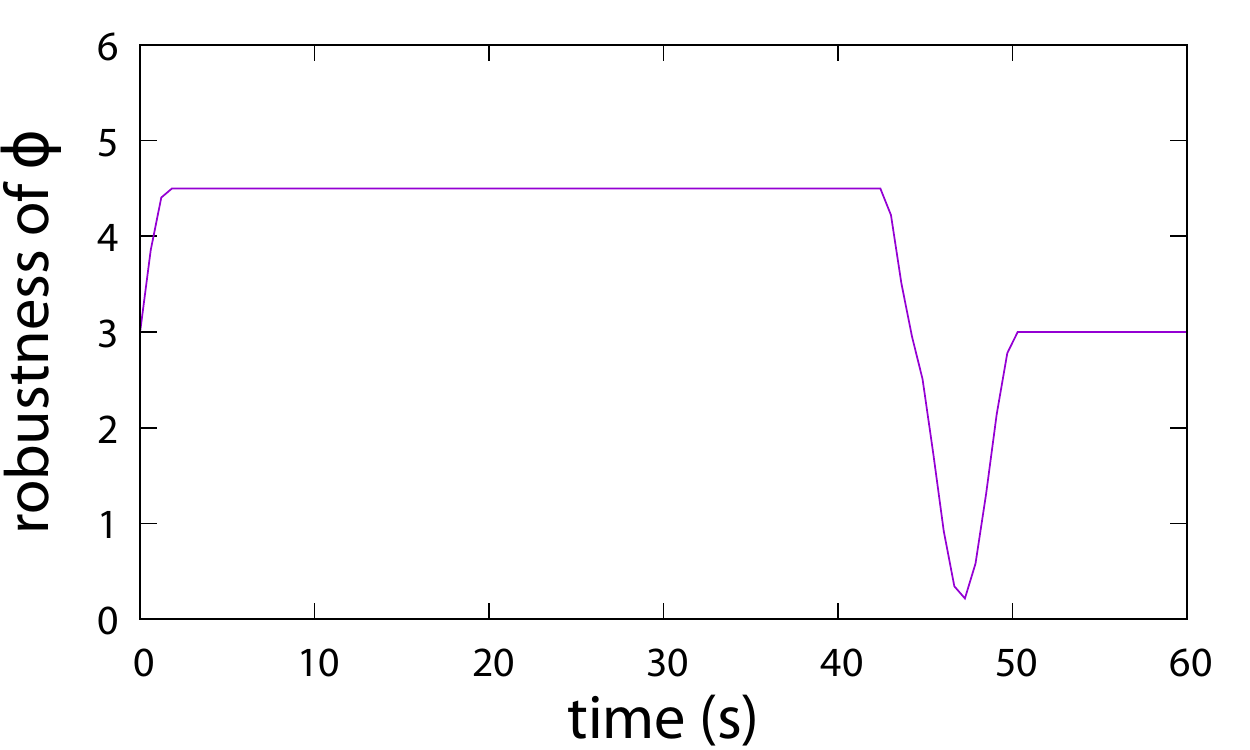}}
\caption{Four graphs showing the value of a hypothetical property $\phi$ over time. The different definitions of robustness assign a different robustness to $\Box \phi$.}
\label{fig:vbools}
\end{center}
\end{figure*}

\subsubsection{\red{Example of max and additive semantics}}

Figure~\ref{fig:vbools} illustrates why sensitivity is important to
falsification. Suppose that the formula to be falsified is $\Box \phi$, and that
this formula happened to be true in the current test case.
Figure~\ref{fig:vbools}(a) illustrates how the robustness of $\phi$ varies with
time in this hypothetical test case. Recall that $\Box \phi$ is computed by
sampling $\phi$ at each time step and taking the conjunction of each sample, up
to a constant factor depending on $\delta t$. In this case, the robustness dips
from $3$ to $1$ at about $t = 48s$, and in both semantics, the robustness of
$\Box \phi$ will be lower compared to if the robustness had been a constant $3$.

Now suppose that the optimizer modifies the test case and observes the output
seen in Figure~\ref{fig:vbools}(b). It seems that this test case is closer to
failing than Figure~\ref{fig:vbools}(a), because there is an extra dip in
robustness. Therefore, we would like the optimiser to prefer (b) to (a), and for
this to happen the robustness of $\Box\phi$ must be lower under (b) than (a).
Under the additive semantics, this is indeed the case, because of sensitivity.
Under the max semantics, however, Figures~\ref{fig:vbools}(a) and (b) give
the same robustness for $\Box \phi$, as the minimum robustness is the same in
both cases. Thus the optimiser is not able to see that moving from
Figure~\ref{fig:vbools}(a) to \ref{fig:vbools}(b) is a good idea. Because the
max semantics is not sensitive, the parameter optimizer is only able to notice changes in the minimum value of $\phi$.

It is not always the case that additive semantics is better than max semantics.
Suppose instead that the optimizer observes the result in
Figure~\ref{fig:vbools}(c). This test case appears much closer to failing than
Figure~\ref{fig:vbools}(a): the minimum is very close to 0. However, the additive
semantics will assign Figure~\ref{fig:vbools}(c) a \emph{higher} robustness than
Figure~\ref{fig:vbools}(a), because the initial segment of the test case has a higher robustness and continues for a long time, which cancels out the lower minimum. The max semantics considers \ref{fig:vbools}(c) to have lower robustness than \ref{fig:vbools}(a), as we might hope.

This problem only occurs because the robustness of the initial segment of the
test case is quite large. Figure~\ref{fig:vbools}(d) shows a less extreme
variant. Both the additive and the max semantics judge this test case as having lower robustness than Figure~\ref{fig:vbools}(a). This is because, if we take two true VBools $(\top, x)$ and $(\top, y)$, their conjunction under the additive semantics is $(\top, z)$ where $z = 1/(1/x+1/y)$. Now we can observe that if $x \ll y$, then $\frac{1}{x} \gg \frac{1}{y}$, so $z = 1/(\frac{1}{x} + \frac{1}{y}) \approx x$. That is, when taking the conjunction of a set of formulas, formulas that have a low robustness have a disproportionate effect on the result. In particular, in the formula $\Box \varphi$, a small decrease in the minimum value cancels out quite a large increase in the maximum value. \red{We also note that $1/(1/x+1/y)$ is always less than $min(x,y)$, \emph{i.e.}, the additive robustness of a conjunction between two true VBools is always smaller than the max robustness. However, as there is no meaning in explicitly comparing the additive robustness value to the max, this does not affect our choice of robustness in any way. }

\begin{figure}[!t]
\begin{center}
  \subfloat[The robustness of $x \land_+ y$.]{\includegraphics[width=0.45\columnwidth]{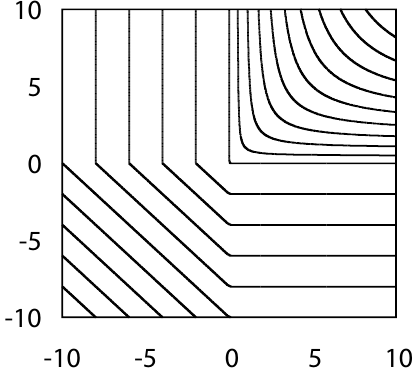}}
  \hskip0.1\columnwidth
  \subfloat[The robustness of $x \land_{max} y$.]{\includegraphics[width=0.45\columnwidth]{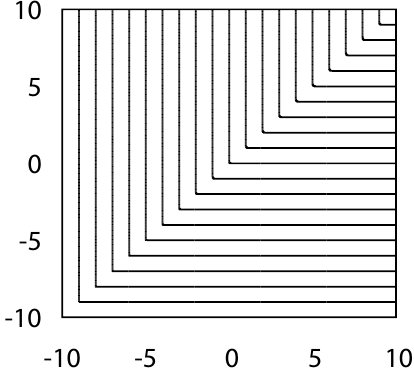}}
\end{center}
\caption{Isobar plots of the robustness of the two semantics of $\land$. Here, negative robustnesses represent false VBools.}
\label{fig:isobars}
\end{figure}

Figure \ref{fig:isobars} illustrates the robustness of $p \land_+ q$ and $p
\land_{max} q$. The $x$-axis gives the robustness of $p$ and the $y$-axis gives
the robustness of $q$; negative values here stand for false VBools. The graph
illustrates the robustness of $p \land q$ using isolines, which connect points
that have equal robustness. Where an isoline is vertical or horizontal, the
connective is insensitive: only changes in a particular argument have an effect
on robustness. We see in the upper-right quadrant that when $p$ and $q$ have
very different robustnesses, $p \land_+ q$ assigns much more importance to the
lower robustness (it starts to approximate the max semantics), but that it
always remains sensitive. This weighting is a deliberate feature of the additive semantics: a subformula with low robustness is likely to be a better target for optimization than a subformula with high robustness, as it it more likely to be easily falsifiable.

\subsection{\red{Other properties of VBools}}

Apart from monotonicity and sensitivity, there are several more commonplace
properties that we would like our semantics to have. The most essential is
\emph{soundness}: a Valued Boolean formula (e.g. $p \land_+ (\lnot_V q \lor+
r)$) and the corresponding Boolean formula (in this case, $p \land (\lnot q \lor
r)$ should always evaluate to the same Boolean result; the only difference is
that the Valued Boolean also computes a robustness. All of the connectives we
have defined are easily seen to be sound, since the Boolean part of each
definition uses the corresponding Boolean connective. Therefore, the choice of
semantics only affects the optimization process, not the truth or falsehood of
the property.

We would also like the usual laws of Boolean logic to hold: connectives should
be associative, commutative, idempotent, have an identity element, have a zero
element, and obey the usual distributivity and negation laws. As mentioned
above, these laws all hold if one ignores the computed robustness, but we would
like robustness to respect these laws too. These properties are important
because we do not want the robustness of a formula to depend on, for example,
how conjunctions are bracketed or what order they are written in, and we do not
want the tester to have to think about what arrangement of brackets is most
suitable.

\red{The max semantics obeys many laws of Boolean logic. Conjunction and disjunction are associative, commutative, idempotent and have $\top_{\!\mathrm{v}}$ and $\bot_{\!\mathrm{v}}$ as identity and zero elements. Distributivity holds, as do De Morgan's laws. What fails are the laws $p
\lor_{max} \lnot_v p = \top$ and $p \land_{max} \lnot_v p = \bot$, since the left and right hand sides may have different robustnesses. Proofs of these laws for VBools are omitted due to space constraints, but they are straightforward and only require an exhaustive case analysis on whether each VBool is true or false.}

The additive connectives satisfy fewer laws than the max semantics. They are
associative, commutative, have $\top_{\!\mathrm{v}}$ and $\bot_{\!\mathrm{v}}$ as identity and zero
elements, and respect de Morgan's laws. They do not satisfy idempotence or
distributivity. Idempotence fails because, for example, $p \land_+ p$ is a
Valued Boolean whose robustness is either twice that of $p$ (if $p$ is false) or
half that of $p$ (if $p$ is true). Distributivity fails for a similar reason,
because expanding $p \land_+ (q \lor_+ r)$ duplicates $p$, increasing its
influence on the robustness computation. We are not aware of a semantics that
combines associativity, commutativity, idempotence and sensitivity; we
conjecture that these four properties are incompatible.\footnote{One could for
example recover idempotence by multiplying or dividing by 2 in the definition of
$\land_+$, but this would destroy associativity.}

To summarise, both max and additive semantics satisfy many Boolean properties, but max satisfies more; in return for giving up some properties, the additive semantics gains sensitivity, which is useful for falsification. In an additive conjunction, the parameter optimizer is able to see when any of the conjuncts' robustness decreases, which is not the case for the max semantics. A final observation is that the additive semantics for conjunction assigns greater weight to less robust conjuncts, which means that when a conjunct is close to being falsified it can be reduced even if this causes the robustness of other conjuncts to increase markedly.

%% file: Sections/5_experiments.tex
\section{Results and Discussion}\label{sec:experiments}
To show the performance of using additive semantics for STL during falsification (compared to max semantics), we perform falsification with additive semantics for four examples. \red{Each of the four examples comes from (or is inspired by) other work, so we refer the reader to these original works for further details about each model.} The results are presented in a set of tables, and the layout of each table is the same. The results are shown in Sections \ref{sec:autotrans} - \ref{sec:static_switched_system}. 

The rows of the tables show which specification is attempted to be falsified, which parameters or specific settings are used, and also which semantics are used. We use the max and additive semantics defined earlier in the paper, but we also include a third \emph{constant} semantics. The robustness value for a constant semantics is equal to 100 if the specification is true, and -100 if the specification is not true. This constant semantics is used as a baseline to verify whether max and additive semantics yield better results than purely random testing\footnote{We have also implemented a \emph{random} semantics, where the robustness at each sample is a uniform random number, but with correct sign (for STL robustness). Falsification for ran dom semantics performs worse than max, additive and constant semantics for all the examples in this paper.}. 

For each parameter setting, the ``Succ'' column shows how many times the specification was actually falsified, and the ``Iter'' column shows the average number of iterations used by the optimization solver in each falsification attempt (the maximum is set to 1000). The ``Iter/Succ'' column shows the average number of iterations for the \emph{falsification attempts that were successful}. The optimization solver used in these examples is a Simulated Annealing solver \cite{abbas2013probabilistic}. 

\subsection{Automatic Transmission Benchmark}\label{sec:autotrans}
The model takes as input the throttle and brake of a vehicle, and simulates the automatic transmission system (for details, see \cite{bardh2014benchmarks}). The model has been used in several other works \cite{jin2015mining, akazaki2015time}, and in this work we perform falsification with the Breach toolbox. The outputs of the system are the vehicle speed ($v$), the engine speed ($\omega$), and the gear. \red{The model contains 69 blocks in total.}

The model is simulated with a fixed-step setting (automatic step size), using the MATLAB solver ode5 (Dormand-Prince).

\subsubsection{Falsification parameters}
The throttle is generated using 7 control points distributed evenly in time, interpolated using the MATLAB interpolation setting \texttt{pchip}. Each control point has a value in the range $[0, 100]$. The brake input is interpolated similarly but only using 3 control points, each in the range $[0, 500]$. 

The specifications to falsify are shown in Table \ref{tab:transmission_specs}. Specifications $\varphi_1$--$\varphi_6$ are taken from \cite{akazaki2015time}. Note, however, that we do not modify the specifications to improve the falsification capability of our additive semantics.

\begin{table}[!t]
\renewcommand{\arraystretch}{1.3}
\caption{Specifications to Falsify for the Automatic Transmission Benchmark. }
\label{tab:transmission_specs}
\centering
\input{tables/transmission_specs.tex}
\end{table}

The comparison between max semantics and additive semantics for each specification is shown in Table \ref{tab:transmission_results}. 

\begin{table*}[!t]
\renewcommand{\arraystretch}{1.1}
\caption{Results for the automatic transmission benchmark. }
\label{tab:transmission_results}
\centering
\input{tables/transmission_results_with_constant.tex}
\end{table*}

\subsection{Abstract Fuel Control Benchmark}\label{sec:AFC}
The model is an Abstract Fuel Control system implemented in Simulink, and it has been proposed as a benchmark for temporal logic falsification \cite{jin2014powertrain}. The inputs to the model are the input throttle $\theta$ (in degrees) and the engine speed $\omega$. The outputs of interest are the Air/Fuel ratio $\lambda$ and the controller mode (either closed-loop or open-loop). The reference value $\lambda_{ref}$ is equal to $14.7$ for the specifications we are considering. \red{The model contains 253 blocks in total.}

The model is simulated with a variable-step setting using the MATLAB solver ode15s (stiff/NDF).

\subsubsection{Falsification parameters}
The engine speed is constant and allowed to be in the range $[900, 1100]$. The throttle angle is generated as a pulse signal with a base value of $8.9$, a delay of $3$, a period in the range $[10, 30]$ and amplitude in the range $[0.1 61]$. Thus, the throttle angle always has a value in the range $[8.9, 69.9]$, always switching back and forth between two values at different times of each simulation. We always simulate the system for 40 seconds. 

The specifications to falsify are shown in Table \ref{tab:AFC_specs}. The specifications are variations of Req. (26) and (27) in \cite{jin2014powertrain}, using $\eta = 1$. 

\begin{table}[!t]
\renewcommand{\arraystretch}{1.3}
\caption{Specifications to Falsify for the Abstract Fuel Control benchmark.}
\label{tab:AFC_specs}
\centering
\input{tables/AFC_specs.tex}
\end{table}

The results for the Abstract Fuel Control benchmark are shown in Table \ref{tab:AFC_results}. 

\begin{table*}[!t]
\renewcommand{\arraystretch}{1.1}
\caption{Results for the Abstract Fuel Control benchmark. }
\label{tab:AFC_results}
\centering
\input{tables/AFC_results_with_constant.tex}
\end{table*}

\subsection{Third Order $\Delta - \Sigma$ Modulator}\label{sec:modulator}
The third order $\Delta - \Sigma$ modulator is used as a technique for analog to digital conversion. The model is described in detail in \cite{dang2004verification} and has previously been used for falsification benchmark purposes \cite{abbas2013probabilistic, aerts2018temporal}. The model has one input $U$, three states $x_1, x_2, x_3$, and three initial conditions $x_1^{init}, x_2^{init}, x_3^{init}$. \red{The model contains 27 blocks in total.}

The model is simulated with a fixed-step setting (automatic step size), using the MATLAB solver discrete (no continuous states).

\subsubsection{Falsification parameters and specification}
The input $U$ is constant during the whole simulation, and the allowed values are in different sets for different scenarios (see Table \ref{tab:modulator_results} for detailed scenarios). The initial conditions are all in the range $[-0.1, 0.1]$. The specifications to falsify are shown in Table \ref{tab:modulator_specs} (note that $\varphi_1^{\Delta - \Sigma}$ is equivalent to $((\varphi_2^{\Delta-\Sigma} \land \varphi_3^{\Delta-\Sigma}) \land \varphi_4^{\Delta-\Sigma})$).

\begin{table}[!t]
\renewcommand{\arraystretch}{1.3}
\caption{Specifications to Falsify for the Third Order $\Delta - \Sigma$ Modulator.}
\label{tab:modulator_specs}
\centering
\input{tables/modulator_specs.tex}
\end{table}

The results for the modulator benchmark are shown in Table \ref{tab:modulator_results}. 

\begin{table*}[!t]
\renewcommand{\arraystretch}{1.1}
\caption{Results for the Third Order $\Delta - \Sigma$ modulator. }
\label{tab:modulator_results}
\centering
\input{tables/modulator_results_with_constant.tex}
\end{table*}

\subsection{Static Switched System}\label{sec:static_switched_system}
The static switched system has no dynamics and is included to show that both max and additive semantics can worsen the performance of falsification, compared to falsifying with Boolean semantics. The model is inspired by \cite{dokhanchi2015requirements}, and it has two inputs $(u_1, u_2) \in [0, 1]^2$ which are kept constant. \red{The model contains 16 blocks in total.} The output $y(t)$ is assigned according to

\begin{equation*}
    y = \begin{cases}
        -2(u_1 + u_2) - 5 & \mathrm{ if } \ \ u_i \geq thresh, \ \forall i \\
        2((u_1 + 1)^2 + (u_2 + 1)^2) & \mathrm{ otherwise.}
  \end{cases}
\end{equation*}

The specification to falsify is $\varphi^{SS} = \Box(y \geq 0)$. In other words, the falsification problem consists of finding a scenario where both inputs have a value above $thresh$. This is difficult since the gradient of the robustness (for max and additive) with respect to the input parameters will point away from the area where the specification is falsified. The results for the static switched system are shown in Table \ref{tab:switched_results}. 

\begin{table*}[!t]
\renewcommand{\arraystretch}{1.1}
\caption{Results for the Static Switched System. }
\label{tab:switched_results}
\centering
\input{tables/switched_results_with_constant.tex}
\end{table*}

\subsection{Transforming Volvo requirements to STL}
We have successfully implemented the framework presented in this paper for transforming causal signal-based specifications into STL. We have transformed the requirements for two industrial models at Volvo Car Corporation, which model the electric machine of an electric vehicle, as well as the battery for an electric vehicle. The models contain $19846$ and $18294$ blocks, respectively\footnote{The block counts include blocks in referenced models.}. \red{Note that the specifications have not been translated manually, only automatically using the procedure presented in this paper. For all the automatically translated specifications, correctness has been asserted during falsification runs, \emph{i.e.}, the Boolean satisfaction of the translated formulas have coincided with the signal value of the specification modeled in Simulink. However, a formal proof of the correctness is out of the scope of this paper, and is considered future work. }

In total, there are 58 transformed requirements for the first model and 36 transformed requirements for the second model. The transformation of requirements into STL specifications have enabled the use of temporal logic falsification for both models. Falsification is now being run continuously for both models, in order to catch software defects during development. Statistics for the transformed STL formulas are shown in Table \ref{tab:STL_stats}. \red{The sheer number of requirements combined with the complexity of the formulas shown in the table indicate that writing the specifications in STL manually would be very time-consuming. }

\begin{table*}[!t]
    \renewcommand{\arraystretch}{1.3}
    \centering
    \caption{Statistics of STL Formulas for the two Volvo models.}
    \begin{tabular}{|c|c|c|c||c|c|c||c|c|c|}
    \hline
        \multirow{2}{*}{Model} & \multicolumn{3}{c||}{Number of operators} & \multicolumn{3}{c||}{Depth} & \multicolumn{3}{c|}{Modal depth} \\
        \cline{2-10}
         & Min & Mean & Max & Min & Mean & Max & Min & Mean & Max \\
        \hline
        Electric machine & 1 & 61.1 & 336 & 1 & 7.65 & 15 & 1 & 2.02 & 4\\
        Battery & 2 & 33.5 & 171 & 1 & 7.95 & 16 & 1 & 2.08 & 4\\
        \hline
    \end{tabular}
    \label{tab:STL_stats}
\end{table*}

\subsection{Discussion}
For the specifications shown in this paper, we can see that no specific semantics perform better than the other two for all models. For several specifications, the constant semantics performs just as well as one or other of the robust semantics.

It is clear from the tables that sometimes max semantics are preferable, and sometimes additive semantics are preferable. For example, for specifications $\varphi_1, \varphi_2, \varphi_7, \varphi_8$ additive semantics clearly perform better, while for specifications $\varphi_5, \varphi_6, \varphi_1^{AFC}, \varphi_2^{AFC}$ max semantics clearly perform better. The static switched system was also introduced to show that the constant semantics (\emph{i.e.} random testing) can be better than max or additive semantics. It is clear that $\varphi^{SS}$ is easier to falsify for the constant semantics than for the other semantics.

Whether a specific semantics outperforms the others depends not only on the specification, but also on the system that is being falsified. 

\subsubsection{Preferable semantics for different specifications}
The intuitive explanation for why additive semantics can be better in some cases is that it takes into account all the different subformulas of $\land$ and $\lor$ formulas (and by extension also the temporal operators). In a conjunction, if only the highest robustness value decreases in between simulations, it is not certain that the max semantics will capture the change, but it will affect the total additive robustness. On the other hand, if one robustness increases while the other robustness decreases, the additive semantics robustness may not be affected, while the max semantics robustness will. 

An example of when it is preferable to notice changes in all clauses of a conjunction is $\varphi_7$. Here, each clause is not difficult to falsify individually, but to be able to falsify them all at once it helps a great deal to include more detailed robustness information about each clause. As such, having conjunctions with many clauses in a specification can indicate that additive semantics would be preferable for that (sub)-specification. 

\subsubsection{Preferable semantics for different systems}
For some system behaviour, it can be non-beneficial to consider changes in all parts of a conjunction. An example of this is the third order $\Delta - \Sigma$ modulator. The results in Table \ref{tab:modulator_results} indicate that $\varphi_4^{\Delta - \Sigma}$ is by far the easiest sub-specification of $\varphi_1^{\Delta - \Sigma}$ to falsify. Including more detailed robustness information about the other sub-specifications ($\varphi_2^{\Delta - \Sigma}$ and $\varphi_3^{\Delta - \Sigma}$) makes the robustness information from $\varphi_4^{\Delta - \Sigma}$ diluted in a sense, meaning that changing from max to additive semantics will not increase falsification capability.

%% file: tables/transmission_specs.tex
\begin{tabular}{|c|c|c|}
\hline
Specification & Formula\\
\hline
\hline
$\varphi_1$ & $\lozenge_{[0, T]}(\omega \geq 2000)$\\
\hline
$\varphi_2$ & $\Box\lozenge_{[0, T]}(\omega \leq 3500 \lor \omega \geq 4500)$\\
\hline
$\varphi_3$ & $\Box_{[0, T]}(\lnot(gear == 4)) $\\
\hline
$\varphi_4$ & $\lozenge(\Box_{[0, T]}(gear == 3))$\\
\hline
$\varphi_5$ & $\bigwedge_{i = 1, \ldots, 4}\Box( (\lnot(gear == i) \land \lozenge_{[0, \epsilon]}(gear == i)$\\
 & $\implies (\Box_{[\epsilon, T + \epsilon]}(gear == i)))$\\
\hline
$\varphi_6$ & $\Box_{[0, T]}(v \leq 85) \lor \lozenge(\omega \geq 4500)$\\
\hline
$\varphi_7$ & $(\Box_{[0,1]}gear == 1) \land (\Box_{[2,4]}gear == 2) $ \\
 & $\land(\Box_{[5,7]}gear == 3) \land (\Box_{[8,10]}gear == 3) $ \\
  & $\land(\Box_{[12,15]}gear == 2)$ \\
\hline 
$\varphi_8$ & $\Box_{[0,20]}\big((gear==4 \land throttle > 45 $\\
 & $\land throttle < 50) \implies \omega < \bar{\omega}\big)$\\
\hline
\end{tabular}

%% file: tables/transmission_results_with_constant.tex
\begin{tabular}{|cc|c|c|c||c|c|c||c|c|c|}
\hline
Specification & Semantics & \multicolumn{9}{c|}{Parameters} \\
\hline

 & & \multicolumn{3}{c||}{$T = 20$} & \multicolumn{3}{c||}{$T = 30$} & \multicolumn{3}{c|}{$T = 40$} \\
\cline{3-11}
& & Succ & Iter & Iter/Succ  & Succ & Iter & Iter/Succ  & Succ & Iter & Iter/Succ  \\
\cline{3-11}
\multirow{3}{*}{$\varphi_1$} & Max & 20 & 103.1 & 103.1 & 19 & 209.1 & 167.5 & 14 & 500.6 & 286.6 \\
 & Additive & 20 & 80.3 & 80.3 & 20 & 133.2 & 133.2 & 20 & 215.1 & 215.1 \\
 & Constant & 14 & 734.0 & 619.9 & 3 & 930.5 & 536.3 & 0 & 1000.0 & - \\
\hline\hline

 & & \multicolumn{3}{c||}{$T = 10$} \\
\cline{3-5}
& & Succ & Iter & Iter/Succ  \\
\cline{3-5}
\multirow{3}{*}{$\varphi_2$} & Max & 16 & 247.9 & 59.9 \\
 & Additive & 20 & 172.1 & 172.1 \\
 & Constant & 20 & 277.3 & 277.3 \\
\hline\hline

 & & \multicolumn{3}{c||}{$T = 4$} & \multicolumn{3}{c||}{$T = 4.5$} & \multicolumn{3}{c|}{$T = 5$} \\
\cline{3-11}
& & Succ & Iter & Iter/Succ  & Succ & Iter & Iter/Succ  & Succ & Iter & Iter/Succ  \\
\cline{3-11}
\multirow{3}{*}{$\varphi_3$} & Max & 0 & 1000.0 & - & 9 & 796.4 & 547.6 & 17 & 467.8 & 373.9 \\
 & Additive & 0 & 1000.0 & - & 10 & 736.0 & 472.0 & 17 & 532.9 & 450.4 \\
 & Constant & 0 & 1000.0 & - & 11 & 641.9 & 348.8 & 16 & 472.9 & 341.1 \\
\hline\hline

 & & \multicolumn{3}{c||}{$T = 1$} & \multicolumn{3}{c||}{$T = 2$} \\
\cline{3-8}
& & Succ & Iter & Iter/Succ  & Succ & Iter & Iter/Succ  \\
\cline{3-8}
\multirow{3}{*}{$\varphi_4$} & Max & 5 & 852.7 & 410.8 & 20 & 182.0 & 182.0 \\
 & Additive & 6 & 795.2 & 317.3 & 20 & 90.6 & 90.6 \\
 & Constant & 1 & 998.4 & 967.0 & 20 & 160.0 & 160.0 \\
\hline\hline

 & & \multicolumn{3}{c||}{$T = 0.8$} & \multicolumn{3}{c||}{$T = 1$} & \multicolumn{3}{c|}{$T = 2$} \\
\cline{3-11}
& & Succ & Iter & Iter/Succ  & Succ & Iter & Iter/Succ  & Succ & Iter & Iter/Succ  \\
\cline{3-11}
\multirow{3}{*}{$\varphi_5$} & Max & 0 & 1000.0 & - & 12 & 754.4 & 590.7 & 20 & 60.1 & 60.1 \\
 & Additive & 0 & 1000.0 & - & 4 & 864.5 & 322.5 & 20 & 90.7 & 90.7 \\
 & Constant & 0 & 1000.0 & - & 13 & 704.6 & 545.5 & 20 & 64.7 & 64.7 \\
\hline\hline

 & & \multicolumn{3}{c||}{$T = 10$} & \multicolumn{3}{c||}{$T = 12$} \\
\cline{3-8}
& & Succ & Iter & Iter/Succ  & Succ & Iter & Iter/Succ  \\
\cline{3-8}
\multirow{3}{*}{$\varphi_6$} & Max & 9 & 731.4 & 403.0 & 20 & 153.5 & 153.5 \\
 & Additive & 12 & 665.9 & 443.1 & 20 & 182.9 & 182.9 \\
 & Constant & 0 & 1000.0 & - & 4 & 899.1 & 495.5 \\
\cline{1-8}\cline{1-8}

 & & \multicolumn{3}{c||}{$ $} \\
\cline{3-5}
& & Succ & Iter & Iter/Succ  \\
\cline{3-5}
\multirow{3}{*}{$\varphi_7$} & Max & 4 & 905.4 & 527.0 \\
 & Additive & 15 & 493.3 & 324.4 \\
 & Constant & 4 & 836.7 & 183.5 \\
\cline{1-8}\cline{1-8}

 & & \multicolumn{3}{c||}{$\hat\omega = 3000$} & \multicolumn{3}{c||}{$\hat\omega = 3500$} \\
\cline{3-8}
& & Succ & Iter & Iter/Succ  & Succ & Iter & Iter/Succ  \\
\cline{3-8}
\multirow{3}{*}{$\varphi_8$} & Max & 20 & 16.9 & 16.9 & 0 & 1000.0 & - \\
 & Additive & 20 & 20.4 & 20.4 & 19 & 296.1 & 259.1 \\
 & Constant & 20 & 10.1 & 10.1 & 0 & 1000.0 & - \\
\cline{1-8}
\end{tabular}

%% file: tables/AFC_specs.tex
\begin{tabular}{|c|c|c|}
\hline
Specification & Formula\\
\hline
\hline
$\varphi_1^{AFC}$ & $\Box_{[11,40]}(|\frac{\lambda(t) - \lambda_{ref}}{\lambda_{ref}}| < tol)$\\
\hline
$\varphi_2^{AFC}$ & $\Box_{[11,35]}\big( (\theta(t) < \theta(t + 0.01) \lor \theta(t) > \theta(t + 0.01)) $\\
 & $\implies \Box_{[1,5]}(|\frac{\lambda - \lambda_{ref}}{\lambda_{ref}}| < tol) \big)$\\
\hline
\end{tabular}

%% file: tables/AFC_results_with_constant.tex
\begin{tabular}{|cc|c|c|c||c|c|c||c|c|c|}
\hline
Specification & Semantics & \multicolumn{9}{c|}{Parameters} \\
\hline

 & & \multicolumn{3}{c||}{$tol = 0.16$} & \multicolumn{3}{c||}{$tol = 0.17$} & \multicolumn{3}{c|}{$tol = 0.18$} \\
\cline{3-11}
& & Succ & Iter & Iter/Succ  & Succ & Iter & Iter/Succ  & Succ & Iter & Iter/Succ  \\
\cline{3-11}
\multirow{3}{*}{$\varphi_1^{AFC}$} & Max & 20 & 313.2 & 313.2 & 9 & 758.6 & 463.7 & 3 & 934.1 & 560.7 \\
 & Additive & 19 & 589.2 & 567.6 & 7 & 837.8 & 536.6 & 2 & 962.0 & 620.0 \\
 & Constant & 14 & 564.4 & 377.7 & 2 & 967.8 & 678.0 & 1 & 971.1 & 423.0 \\
\hline\hline

 & & \multicolumn{3}{c||}{$tol = 0.16$} & \multicolumn{3}{c||}{$tol = 0.17$} & \multicolumn{3}{c|}{$tol = 0.18$} \\
\cline{3-11}
& & Succ & Iter & Iter/Succ  & Succ & Iter & Iter/Succ  & Succ & Iter & Iter/Succ  \\
\cline{3-11}
\multirow{3}{*}{$\varphi_2^{AFC}$} & Max & 20 & 319.4 & 319.4 & 7 & 786.6 & 390.4 & 2 & 954.8 & 548.0 \\
 & Additive & 19 & 545.5 & 521.6 & 2 & 942.8 & 428.0 & 2 & 962.5 & 625.5 \\
 & Constant & 12 & 680.8 & 468.0 & 3 & 937.8 & 585.0 & 3 & 929.1 & 527.7 \\
\hline
\end{tabular}

%% file: tables/modulator_specs.tex
\begin{tabular}{|c|c|c|}
\hline
Specification & Formula\\
\hline
\hline
$\varphi_1^{\Delta - \Sigma}$ & $\Box\left(\bigwedge_{i=1}^3 (-1 \leq x_i \land x_i \leq 1 )\right).$\\
\hline
$\varphi_2^{\Delta - \Sigma}$ & $\Box\left(-1 \leq x_1 \land x_1 \leq 1\right)$\\
\hline
$\varphi_3^{\Delta - \Sigma}$ & $\Box\left(-1 \leq x_2 \land x_2 \leq 1\right)$\\
\hline
$\varphi_4^{\Delta - \Sigma}$ & $\Box\left(-1 \leq x_3 \land x_3 \leq 1\right)$\\
\hline
\end{tabular}

%% file: tables/modulator_results_with_constant.tex
\begin{tabular}{|cc|c|c|c||c|c|c||c|c|c|}
\hline
Specification & Semantics & \multicolumn{9}{c|}{Parameters} \\
\hline

 & & \multicolumn{3}{c||}{$U \in [-0.35, 0.35]$} & \multicolumn{3}{c||}{$U \in [-0.40, 0.40]$} & \multicolumn{3}{c|}{$U \in [-0.45, 0.45]$} \\
\cline{3-11}
& & Succ & Iter  & Iter/Succ  & Succ & Iter  & Iter/Succ  & Succ & Iter  & Iter/Succ  \\
\cline{3-11}
\multirow{3}{*}{$\varphi_1^{\Delta - \Sigma}$} & Max & 12 & 799.7 & 666.2 & 19 & 281.6 & 243.7 & 20 & 144.3 & 144.3 \\
 & Additive & 20 & 296.8 & 296.8 & 20 & 335.1 & 335.1 & 12 & 730.8 & 551.3 \\
 & Constant & 20 & 205.5 & 205.5 & 17 & 513.3 & 427.4 & 4 & 875.0 & 374.8 \\
\hline\hline

 & & \multicolumn{3}{c||}{$U \in [-0.35, -0.35]$} \\
\cline{3-5}
& & Succ & Iter & Iter/Succ   \\
\cline{3-5}
\multirow{3}{*}{$\varphi_2^{\Delta - \Sigma}$}& Max & 0 & 1000.0 & - \\
 & Additive & 0 & 1000.0 & - \\
 & Constant & 0 & 1000.0 & - \\
\cline{1-5}\cline{1-5}

 & & \multicolumn{3}{c||}{$U \in [-0.35, -0.35]$} \\
\cline{3-5}
& & Succ & Iter & Iter/Succ   \\
\cline{3-5}
\multirow{3}{*}{$\varphi_3^{\Delta - \Sigma}$} & Max & 0 & 1000.0 & - \\
 & Additive & 0 & 1000.0 & - \\
 & Constant & 0 & 1000.0 & - \\
\cline{1-5}\cline{1-5}

 & & \multicolumn{3}{c||}{$U \in [-0.35, -0.35]$} \\
\cline{3-5}
& & Succ & Iter & Iter/Succ   \\
\cline{3-5}
\multirow{3}{*}{$\varphi_4^{\Delta - \Sigma}$} & Max & 13 & 627.0 & 426.2 \\
 & Additive & 12 & 739.5 & 565.9 \\
 & Constant & 5 & 872.6 & 490.4 \\
\cline{1-5}
\end{tabular}

%% file: tables/switched_results_with_constant.tex
\begin{tabular}{|cc|c|c|c||c|c|c||c|c|c|}
\hline
Specification & Semantics & \multicolumn{9}{c|}{Parameters} \\
\hline

 & & \multicolumn{3}{c||}{$thresh = 0.7$} & \multicolumn{3}{c||}{$thresh = 0.8$} & \multicolumn{3}{c|}{$thresh = 0.9$} \\
\cline{3-11}
& & Succ & Iter & Iter/Succ  & Succ & Iter & Iter/Succ  & Succ & Iter & Iter/Succ  \\
\cline{3-11}
\multirow{3}{*}{$\varphi^{SS}$} & Max & 15 & 566.3 & 421.7 & 10 & 741.1 & 482.2 & 3 & 937.7 & 584.7 \\
 & Additive & 16 & 518.4 & 397.9 & 7 & 861.3 & 603.7 & 3 & 943.3 & 622.0 \\
 & Constant & 20 & 118.3 & 118.3 & 20 & 136.1 & 136.1 & 20 & 373.4 & 373.4 \\
\hline
\end{tabular}

%% file: Sections/6_conclusion.tex
\section{Conclusions}\label{sec:conclusions}

We have presented two additions to potentially increase the capability of falsification of temporal logic specification for Cyber-Physical Systems (CPSs). 

The first addition is a specification transformation framework, which takes requirements modeled in a causal signal-based frameworks and transforms them into Signal Temporal Logic (STL) formulas. The framework has been implemented for the specifications in two industrial-sized models at Volvo Car Corporation, and it has enabled the use of falsification for both of the models. The specification transformation outputs a specification where we also have information about which preconditions should be fulfilled for different parts of the specification to be evaluated for given signal values and a given time. 

The second addition is the introduction of additive semantics in the falsification process. Considering the established robust semantics of STL formulas as the max semantics, the difference for additive semantics is that the robustness of each clause in a conjunction can affect the total robustness of the conjunction, even if only one of the clause's robustness changes. Disjunction and temporal operators are defined in terms of conjunction for the additive semantics. 

To indicate the usability of additive semantics for falsification, we have compared them to max semantics as well as \emph{constant} semantics (essentially only Boolean information and no robustness) for several different models and specifications. The models we show results for are both well-known benchmark models, as well as a simple non-dynamic model to prove that all three choices of semantics can be the most viable. Previous work on falsification has overlooked the need to compare against the constant semantics as a baseline; our evaluation made it clear that for some specifications, falsification had no benefit over random testing. \red{However, for most cases excluding the system in Section \ref{sec:static_switched_system}, it is clear that constant semantics perform worse that the others, indicating that robustness-based falsification is a reasonable way of finding faults in CPSs}. We encourage other researchers to include a baseline comparison in their future work.

Which of the three semantics performs best depends both on the specification and the model. In a black-box setting, it is thus very difficult to decide which semantics to use for which operator in the specification to get the best results for falsification. 

\subsection{Future work}

We have so far defined two semantics for Valued Booleans. There are most likely
many more, each with their own trade-offs; we plan to explore these. Also, since
the best choice of semantics can be different for each connective in a given
specification, we would like to both
\begin{itemize}
    \item formulate principles that can guide a tester in choosing a suitable semantics for each operator in a given specification, and 
    \item analyze \emph{both} the model and the specification to reason about which semantics would be best for falsification (\emph{i.e.} grey-box or white-box testing). 
\end{itemize}

\red{Evaluating the effect of different robust semantics in falsification of generated specifications for the industrial examples presented in this paper is also a path we plan to explore. A more theoretical approach is about how the additive robustness relates to the view of temporal logic as filtering.} Finally, it would be interesting to look at falsification which includes the extra information that we get from the specification transformation presented in this paper -- namely, the information about all the preconditions that need to be fulfilled for different parts of the specification to be evaluated.